\pdfoutput=1
\documentclass{iopart}

\usepackage{comment}
\usepackage{graphicx}

\usepackage{iopams} 
\newcommand{\cev}[1]{\reflectbox{\ensuremath{\vec{\reflectbox{\ensuremath{#1}}}}}}
\newcommand{\ix}[1]{\ensuremath{\mathrm{#1}}}

\newcommand{\inv}{\ix{inv}}

\usepackage{amsopn}
\DeclareMathOperator{\sgn}{sgn}

\begin{document}

\title[]{Two-particle irreducible functional renormalization group
  schemes---a comparative study}

\author{J.F. Rentrop, S.G. Jakobs and V. Meden}

\address{Institut f\"ur Theorie der Statistischen Physik, RWTH Aachen
  University and JARA - Fundamentals of Future Information Technology,
  52062 Aachen, Germany}

\ead{jan.rentrop@rwth-aachen.de}

\begin{abstract} 

  We derive functional renormalization group schemes for Fermi systems
  which are based on the two-particle irreducible approach to the
  quantum many-body problem. In a first step, the cutoff is introduced
  in the non-interacting propagator as it is commonly done in
  functional renormalization group based on one-particle irreducible
  vertex functions. The most natural truncation of the resulting
  infinite hierarchy of flow equations is shown to be fully equivalent
  to self-consistent perturbation theory. An earlier suggested
  alternative truncation strategy is considered as well. In a second
  step, the cutoff is introduced in the two-particle interaction. Again
  two truncation procedures are investigated, one of which was derived
  before. In the latter, the mean-field solution of the many-body
  problem is considered as the starting point of the renormalization
  group flow. We compare the performance and the required numerical
  resources for solving the coupled flow equations for all the
  approximate schemes by applying them to the problem of the quantum
  anharmonic oscillator.  In a functional integral representation, this
  model has a formal similarity to the quantum many-body problem. The
  perspectives for applying the derived two-particle irreducible
  functional renormalization group approaches to zero- and
  one-dimensional systems of correlated fermions are discussed.
 
\end{abstract} 

\maketitle

\section{Introduction}

Over the past years the functional renormalization group (fRG) was
established as a versatile tool to study low-dimensional interacting
Fermi systems \cite{Met12,Kopietzbook}. The variant almost exclusively
used up to now is based on the generating functional for the
one-particle irreducible (1PI) vertex functions. In this approach, one
derives an exact infinite hierarchy of coupled differential flow
equations for the 1PI vertex functions (self-energy and effective
$n$-particle interactions), where the derivative is taken with respect
to a flow parameter. Prior to practical calculations for any fermionic
many-body model, a truncation of this set of coupled flow equations is
required.  Most truncation schemes were guided by perturbation theory
in the flowing two-particle interaction rendering 1PI fRG a
non-perturbative approximate method which is controlled for small to
intermediate interactions. It is one of the assets of fRG based
approximation schemes that they are unbiased in the sense that barely
any a posteriori knowledge of the physics of the problem at hand must
be used when setting them up. This e.g. allows for a faithful analysis
of the interplay of distinct emergent quantum many-body phenomena.

The 1PI fRG was used to investigate the competing ordering tendencies (e.g. magnetism and 
unconventional superconductivity) in the two-dimensional Hubbard model and variants of 
the latter \cite{Met12,Pla13}, to study the spectral and transport properties of 
inhomogeneous one-dimensional lattice models (quantum wires) falling into the 
Luttinger liquid universality class \cite{Met12}, as well as of quantum impurity 
(quantum dot) models \cite{Met12}. Compared to other RG approaches, fRG has the 
distinct advantage that it can directly be applied to microscopic lattice models and 
not only to low-energy effective field theories. This way, high-energy features and 
complex crossover behavior are captured. 

For problems of interest in which plain perturbation theory in the two-particle 
interaction of characteristic amplitude $U$ is divergent in the infrared limit 
(e.g. in two-dimensional models or in translationally invariant Luttinger liquids), 
the flow parameter must be introduced such that it has infrared regularizing 
properties; it acts as a low-energy cutoff \cite{Met12}. In other systems, low-order 
perturbation theory is regular but, due to logarithmic terms with prefactor 
$U^n$, $n \in {\mathbb N}$, restricted to parameter regimes of infinitesimal extend. 
Then, 1PI fRG can be used to resum the logarithms generically leading to power-law 
behavior with $U$-dependent exponents \cite{Met12}.
In both cases, characteristics are utilized which are commonly associated to RG 
methods, e.g. the successive inclusion of energy scales from high to low and 
the resummation of logarithmic terms. However, 1PI fRG was also used for problems 
in which perturbation theory is regular and not plagued by logarithms. For such, 
fRG should be viewed as a ``renormalization group enhanced perturbation theory'' 
in which certain types of diagrams are resummed \cite{Met12,Hed04,Kar06,Kar08}. It is then
possible to employ other flow parameter schemes than the introduction of an infrared cutoff. 
One can e.g. think of successively turning on the strength of the two-particle interaction 
during the flow (see e.g. Ref.~\cite{Hon04}).     

Recent attempts to abandon the realm of small to intermediate
interactions within fRG include to set up 1PI schemes in which certain
aspects of the interaction are already included in the initial point
of the RG flow \cite{Kin13,Tar14,Went14}; e.g. the idea of taking
dynamical mean-field theory containing all the local correlations as
the starting point of the fRG flow was put forward \cite{Tar14}. In
another attempt for a certain quantum impurity model, namely the
single-impurity Anderson model, lots of a posteriori knowledge of the
physics was built into the fRG procedure \cite{Str13}. In the latter
case, however, one gives up the strength of fRG of being unbiased.
The 1PI fRG was recently extended to study quantum wires as well as
quantum dots in non-equilibrium. The driven non-equilibrium steady
state \cite{Jak07,Gez07,Jak10b,Kar10} as well as the non-equilibrium
time evolution \cite{Ken12} were investigated.

Despite its successes, present day 1PI fRG suffers from several shortcomings. 
For general two-dimensional models, it still constitutes a challenge to 
formulate an unbiased purely fermionic scheme which allows to extend the  flow 
into phases with spontaneously broken symmetry \cite{Met12}.  Phase transitions are 
driven by composite fields and not the bare fermions. Crudely speaking, these 
composite degrees of freedom are not properly dealt with in the standard 
implementation of 1PI fRG. As a consequence, the RG flow diverges at a certain 
cutoff scale and has to be stopped. Therefore, only ordering tendencies can be 
read off but one cannot compute observables of the cutoff-free problem in the 
symmetry broken state. In recently developed novel schemes the flow is 
initialized including a small selected symmetry breaking term. In the parts of 
the parameter space in which the corresponding instability dominates, the flow 
is regularized and the cutoff can be removed allowing to investigate the 
symmetry broken phase \cite{Sal04,Ger08,Ebe13,Ebe14,Mai14}. Another way out is to introduce the bosonic composite degrees of 
freedom during the RG flow by partial "bosonization" \cite{Kopietzbook,Gie02,Bai04,Die07,Fri11,Bir05,sue06}, however, again to the price of giving up 
treating all instabilities on equal footing.

For zero-dimensional quantum impurity models, the above mentioned
``flow to strong coupling'' is not an issue and the truncated set of
1PI fRG flow equations can be integrated down to vanishing
cutoff. From the resulting expressions for the vertex functions,
approximate results for observables of interest such as the ground
state energy or the single-particle spectral function of the cutoff-free model can be computed.  However, case studies revealed that going
to higher orders in the truncation does not necessarily lead to the
hoped for improvement of the results beyond the limit of weak
interactions (for which the results systematically improve).  E.g. for
the single-impurity Anderson model, the local spin susceptibility
shows Kondo physics in first order truncation \cite{Kar06} but loses
this property in second order \cite{Hed04,Kar08}.  Similarly, for the
interacting resonant level model, expected power-law behavior of
observables confirmed in first order truncation is lost in second
order \cite{Kar10}.

A general drawback of the 1PI fRG for practical computations is that it 
is computationally costly to systematically include frequency dependence. 
This restricts the access to dynamical observables. A frequency dependence of the 
self-energy is only generated if the dependence of the two-particle 
vertex on its three independent frequencies 
(for a time translational invariant model) is kept \cite{Met12}. This leads 
to an excessive growth of the number of equations to be 
considered.\footnote{For an approximate treatment of the frequency dependence, see 
Refs.~\cite{Hed04,Kar08,Jak10b}.}
Therefore, 
it was only for zero-dimensional quantum dot models with a few interacting degrees 
of freedom \cite {Hed04,Kar08} or toy models such as the quantum harmonic 
oscillator with quartic anharmonicity \cite{Hed04} that frequency dependence up 
to second order could be included completely. This e.g. prevented the 
study of bulk Luttinger liquid physics in one-dimensional models of interacting
electrons. Schemes to include a frequency dependent self-energy 
for two-dimensional systems were put forward in Refs. \cite{Hus09,Gie12,Ebe13}. 

Finally, it remains a challenge to fulfill Ward identities in truncated 1PI fRG 
approaches beyond the order in which the considered set of flow equations 
contains all diagrams of plain perturbation theory \cite{Kat04,Ens05,Str13}. 
Therefore, the relation between approximate 1PI fRG schemes and so-called 
conserving approximations  \cite{Bay61,Bay62} remains 
elusive.\footnote{For recent considerations on the relation between fRG and 
Schwinger-Dyson equations, see Ref.~\cite{Ves13}.} 
  
To circumvent the first mentioned drawback of 1PI fRG for problems of
two-dimensional fermions, it was suggested in Refs. \cite{Dup05,Dup14}
to set up fRG schemes based on the two-particle irreducible (2PI)
approach to quantum many-body physics \cite{Bay61,Bay62}.  This would
allow to better incorporate the composite degrees of freedom.  Such a
procedure was earlier hinted at in Ref. \cite{Wet07}.  However, recent
results obtained by dynamical mean-field theory for models with phase transitions
\cite{Sch13,Jan14} reveal unforeseen divergencies in the 2PI
two-particle vertex. These might lead to
obstacles for a 2PI fRG treatment of two-dimensional models.

We are primarily interested in interacting quantum dots and wires and
here study 2PI fRG approaches to investigate their potential to
overcome the second and third problem of 1PI fRG. Furthermore, using
the same 2PI framework for fRG schemes as in standard conserving
approximations promises insights on the relation to these.

In a first step, we derive four pair-wise related but different 2PI
based fRG schemes. Two of them were introduced earlier
\cite{Dup05,Dup14}.  For these, we suggest modifications and
provide insights which were not emphasized so far. Furthermore, we
allude to the relations between all the approximate
approaches. Instead of studying a fermionic many-body problem, we then
apply these methods in a comparative study to the problem of the
quantum harmonic oscillator with quartic anharmonicity.  In a
functional integral representation of the partition function, the
anharmonic oscillator has the same structure as the one of a fermionic
many-body problem (although with real fields instead of Grassmann
fields).  The anharmonicity $g$ corresponds to the amplitude of the
two-particle interaction in a many-body problem.  Considering this toy
model, we thus reduce the numerical effort (due to the reduced number of
degrees of freedom and the more rapid decay of propagators) without
sacrificing the general structure of the equations to be solved. We
emphasize that the formal complexity of the problem we study is
comparable to that of the single-impurity Anderson model \cite{Hed04}.
The anharmonic oscillator was earlier used to illustrate the fRG
procedure and to test newly developed fRG schemes
\cite{Hed04,Aok02,Gie06,Wey06,Nag11}. It has the additional advantage that
numerically exact results for observables of interest can easily be
obtained. By comparing exact results to our approximate ones for
the dependence of the ground state energy and the position
fluctuations $\left< x^2 \right>$ on $g$ as well as for the frequency
dependence of the self-energy at fixed $g$, we obtain an impression of
the approximation quality.  The 2PI fRG results are in addition
compared to the ones of 1PI fRG.  We furthermore discuss the
advantages and drawbacks of the different schemes in their application
to fermionic many-body problems.

We briefly study another fRG scheme which is not based on a 2PI
generating functional but on one which is two-particle
point-irreducible (2PPI). The use of this was suggested \cite{Pol02,swe04}
for its relation to density functional theory and applied to the
anharmonic oscillator in Ref. \cite{Kem13}. The results of the 2PPI
approach to the anharmonic oscillator are compared to those of 2PI.

We identify two 2PI schemes in which a frequency dependent
self-energy is generated consistently without having to deal with
functions of three independent frequencies. Compared to the standard
frequency dependent 1PI approach this considerably reduces the
numerical effort. Nevertheless, the accuracy of these two 2PI schemes
for the anharmonic oscillator is satisfactory. For one of them, it is
even slightly superior to 1PI fRG. It remains to be seen if the same
holds for fermionic many-body systems.  We furthermore prove that the
most natural truncation of the 2PI flow equations with the flow
parameter introduced via the free single-particle propagator exactly
gives the well known self-consistent perturbation theory (conserving
approximation), e.g. in lowest order the self-consistent Hartree-Fock
approximation.

Our paper is organized as follows. In Sect.~\ref{sec:part}, we give the
functional integral representation of the partition function of
interest. The generating functionals of the 2PI formalism are
introduced in Sect.~\ref{sec:gen-func}. In Sect.~\ref{sec:C-flow}, we
present a first 2PI fRG approach which results from including the flow
parameter in the free propagator $C$ (as is mostly done in 1PI fRG);
we coin it $C$-flow.  The two approximation schemes resulting from
$C$-flow as well as their application to the anharmonic oscillator are
presented in subsections.  As a sequel of Sect.~\ref{sec:gen-func}, we present technical details of the
pair propagator and the Bethe-Salpeter equation in
Sect.~\ref{sec:Bethe-Salpeter} which are crucial for
the second 2PI fRG approach, the so-called $U$-flow. In this approach, the flow
parameter is introduced via the amplitude $U$ of the two-particle
interaction as described in Sect.~\ref{sec:U-flow}.  Again, subsections
are devoted to two resulting approximation schemes.  In one, the
non-interacting problem is considered as the starting point of the RG
flow; in the other, the flow is started from the mean-field solution of
the problem at hand. The results of these schemes for the anharmonic
oscillator are presented in another subsection. In
Sect.~\ref{sec:DFT-fRG}, we give a brief account of the 2PPI fRG and
discuss our results for the anharmonic oscillator.  The results for
the most successful approximate 2PI and 2PPI schemes are compared to
the ones of 1PI in a concluding Sect.~\ref{sec:comp_schemes}. It also
contains a brief discussion of the perspectives of applying 2PI fRG to
fermionic quantum many-body problems. In the Appendix, we give some
details of our numerical implementation of the frequency dependence.


\section{The fermionic many-body problem and the anharmonic oscillator}
\label{sec:part}

We are interested in a 2PI fRG approach to fermionic quantum many body
problems. The starting point for the definition of suitable generating
functionals is the grand canonical partition function of a system of
interacting fermions written in coherent state functional integral
representation \cite{Neg87},
\begin{equation}
  Z = \tr e^{-\beta(H-\mu N)} = \int D[\psi] e^{-S[\psi]} ,
\end{equation}
where the action is
\begin{equation}
  \label{eq:action}
  \fl
  S[\psi]
  = 
  -\frac{1}{2} \sum_{\alpha \alpha'} \psi_{\alpha}
  \left(C^{-1}\right)_{\alpha \alpha'} \psi_{\alpha'}  
  +
  \frac{1}{4!}\! \sum_{\alpha_1 \alpha'_1 \alpha_2 \alpha'_2} \!
  U_{\alpha_1 \alpha'_1 \alpha_2 \alpha'_2} \psi_{\alpha_1}
  \psi_{\alpha'_1} \psi_{\alpha_2}\psi_{\alpha'_2}. 
\end{equation}
Following Refs.~\cite{Dup05, Dup14}, we use a fully antisymmetrized
notation here. This is customary in 2PI approaches and allows to
derive the Bethe-Salpeter equation in arbitrary channels, compare
Sect.~\ref{sec:Bethe-Salpeter}. The notation is based on indices
$\alpha=(c,\tau, y)$ with charge index $c=\pm$, imaginary time $\tau$
and single-particle quantum number $y$. The fields $\psi_{-,\tau
  y}=\psi_y(\tau)$, $\psi_{+,\tau y} = \overline \psi_y(\tau)$ take
Grassmann number values. The inverse free propagator is antisymmetric,
$C^{-1}_{\alpha'\alpha}=-C^{-1}_{\alpha\alpha'}$, and satisfies
\begin{equation}
  C^{-1}_{(+,\tau_1 y_1)(-,\tau_2 y_2)} 
  =
  -\delta(\tau_1-\tau_2)(\partial_{\tau_2}+\epsilon_{y_1 y_2}),
\end{equation}
where $\epsilon$ is the matrix of $(H_0-\mu N)$ in the single-particle
basis. Here, $H_0$ denotes the non-interacting part of the Hamiltonian $H$. The two-particle interaction vertex $U$ is fully antisymmetric
under permutation of its indices, $U_{\alpha_1 \alpha_2 \alpha_3
  \alpha_4} = (-1)^P U_{\alpha_{P1} \alpha_{P2} \alpha_{P3}
  \alpha_{P4}}$.

In the following, we study various 2PI fRG methods.  Since the
application of such schemes to many-body problems (like the Anderson
impurity model) is involved, we here employ them to a simpler toy model
in a first step. For that purpose, a ``classical'' choice
\cite{Hed04,Gie06,Wey06,Nag11} is the quantum anharmonic oscillator
with Hamiltonian
\begin{equation}
  H
  =
  \frac{\hat{p}^2}{2} + V(\hat{x})
  =
  \frac{\hat{p}^2}{2} + \frac{\omega^2_\ix{G}}{2}\hat{x}^2 + \frac{g}{4!} \hat{x}^4.
\end{equation}
The position eigenstate path integral expression for its canonical partition function is
\begin{equation}
   Z = \tr e^{-\beta H} = \int D[x] e^{-S[x]}
\end{equation}
with real fields $x$, a suitably normalized measure $D[x]$, and the corresponding action (see e.g. Ref.~\cite{Neg87})
\begin{equation}
 \fl S = \!
  \int_0^\beta \!\!d\tau \left\{ \frac{\left[\partial_\tau x(\tau)\right]^2}{2} \!+\! V\big(x(\tau)\big) \right\}
  = \! \int_0^\beta \!\!d\tau \left\{ \frac{1}{2}
    x(\tau) \left(\cev{\partial}_\tau \vec{\partial}_\tau \!+\! \omega_\ix{G}^2 \right)  x(\tau)
  \!+\! \frac{g}{4!} \left[x(\tau)\right]^4 \right\}.
\end{equation}
Here, $\cev{\partial}_\tau$ ($\vec{\partial}_\tau$) denotes a partial derivative that acts to the left (right). If we define the totally symmetric quantities
\begin{eqnarray}
  C^{-1}(\tau,\tau') =
  - \cev{\partial}_{\tau} \delta(\tau-\tau')
    \vec{\partial}_{\tau'} - \delta(\tau-\tau') \omega_\ix{G}^2,
  \\
  U(\tau_1, \tau'_1,\tau_2, \tau'_2)  =
  \delta(\tau_1^\prime-\tau_1)\delta(\tau_2-\tau_1)\delta(\tau'_2-\tau_1)g,
\end{eqnarray}
and identify $\tau=\alpha$ as well as $x=\psi$, the action has
formally the same structure as given in Eq.~(\ref{eq:action}) for the
many-body problem. Due to the missing charge and state indices,
computations for the quantum anharmonic oscillator are analytically
simpler and numerically faster than for actual many-body
problems. Typical observables can even be computed numerically exact
which allows to assess the quality of approximate methods. Of course,
the results cannot be directly transferred to many-body problems,
since the fields are of different nature (real fields vs. Grassmann
ones) and the additional degrees of freedom in many-body problems
often lead to much richer physics. Nevertheless, basic properties of
different approximation procedures can be tested on the anharmonic
oscillator.

The general derivations in the following sections hold for the fermionic many-body problem as well as for the anharmonic oscillator. Occasionally, the different commutation properties of Grassmann and real numbers lead to different sign prefactors. Then, $\zeta=-1$ applies to fermions and $\zeta=+1$ to the anharmonic oscillator. 


\section{Generating functionals in the 2PI formalism}
\label{sec:gen-func}

In this section, we briefly recall some basic relations for the
generating functionals in the 2PI formalism and introduce the
corresponding notation.  We start by defining the functional
\begin{equation}
 \label{eq:Z_of_J}
 Z[J] = 
 e^{W[J]} =
 \int\! D[\psi] e^{-S[\psi] + \frac{1}{2} \sum_{\alpha
     \alpha'} 
   \psi_{\alpha} J_{\alpha \alpha'} \psi_{\alpha'}}.
\end{equation}
The external source $J$ is coupled to two fields here
(cf. Refs.~\cite{Neg87,Paw07}),
which is decisive to obtain 2PI quantities by the Legendre
transformation given below. $J$ is defined to have the same
(anti-)symmetry as $C^{-1}$, that is $J_{\alpha' \alpha}=\zeta
J_{\alpha \alpha'}$. The first functional derivative of $W[J]$ leads
to the (time ordered) expectation value of $\psi_\alpha
\psi_{\alpha'}$, that is to the propagator,
\begin{equation}
  W^{(1)}_{\alpha\alpha'}[J] = 
  \frac{\delta W[J]}{\delta J_{\alpha \alpha'}} =
  \langle \psi_\alpha \psi_{\alpha'} \rangle_J
  =-G_{\alpha \alpha'}[J].
\end{equation}
In the following, it turns out to be convenient to use a combined
index $\gamma=(\alpha,\alpha')$. The higher derivatives
\begin{equation}
  W^{(n)}_{\gamma_1 \dots \gamma_n} 
  = \frac{\delta^n W}{\delta J_{\gamma_1} \dots \delta J_{\gamma_n}}
\end{equation}
are the $2n$-point Green functions that are connected if each pair $(\alpha_i,\alpha'_i)$ is considered intrinsically connected.  They obey the symmetry relations
\begin{equation}
  \label{eq:index-perm-symm}
  W^{(n)}_{\gamma_1 \dots \gamma_n} = W^{(n)}_{\gamma_{P1} \dots
    \gamma_{Pn}},
  \qquad
  W^{(n)}_{(\alpha_1, \alpha'_1),\gamma_2 \dots \gamma_n} = \zeta
  W^{(n)}_{(\alpha'_1, \alpha_1),\gamma_2 \dots \gamma_n}
\end{equation}
for arbitrary permutations $P$.

The Legendre transformation of $W[J]$ leads to a functional of the free variable $G$ which is called 2PI effective action,
\begin{equation}
 \Gamma[G]=\left.\left\{-W[J]-J \cdot G\right\}\right|_{J[G]}.
\end{equation}
Here we used a dot-product notation,
\begin{equation}
  J \cdot G 
  = \frac{1}{2} \sum_\gamma J_\gamma G_\gamma 
  = \frac{1}{2} \sum_{\alpha \alpha'} J_{\alpha \alpha'} G_{\alpha
    \alpha'}
  = \frac{\zeta}{2} \tr JG
  = G \cdot J, 
\end{equation}
where the factor $\frac{1}{2}$ is related to the (anti-)symmetry of the factors under exchange of $\alpha$ and $\alpha'$. As the quadratic part of the action is given by $(C^{-1} + J)\cdot (\psi \psi)$, it follows that $\Gamma[G]=\Gamma[G,C]$ depends explicitly on $C$, namely due to an addend $C^{-1} \cdot G = \frac{\zeta}{2}\tr C^{-1}G$.  In the non-interacting case, one finds
\begin{eqnarray}
  W_0[J]
  = \ln [\det(-C^{-1} - J)]^{-\zeta/2}
  = -\frac{\zeta}{2} \tr \ln (-C^{-1}-J),
  \\
 \Gamma_0[G] 
 =-\frac{\zeta}{2} \big[ \tr\ln(-G) - \tr (C^{-1} G - 1)\big].
\end{eqnarray}
The Luttinger-Ward functional $\Phi$ is defined as
\begin{equation}
  \Phi[G]=\Gamma[G]-\Gamma_0[G]
\end{equation}
and does not depend on $C$. The first functional derivative of $\Gamma$ and $\Phi$ is
\begin{eqnarray}
  \Gamma^{(1)}_{\gamma}[G] = \frac{\delta \Gamma[G]}{\delta G_\gamma} 
  = -J_{\gamma}[G],
  \\
  \Phi^{(1)}_{\gamma}[G] = \frac{\delta \Phi[G]}{\delta G_\gamma} =
  G^{-1}_\gamma - C^{-1}_\gamma  -J_{\gamma}[G] = - \Sigma_\gamma[G],
\end{eqnarray}
where $\Sigma$ denotes the self-energy. In the last step, we used that
the effective free (i.e. non-interacting) propagator in the model with
external sources is given by $(C^{-1}+J)^{-1}$. As $\Sigma[G]$ is the
sum of all amputated one-particle 2PI diagrams with full propagator
lines, it follows that $\Phi[G]$ is the sum of all vacuum 2PI diagrams
(which are all 3PI) with full propagator lines.
\begin{figure}
  \begin{center}
    \includegraphics[width=\textwidth]{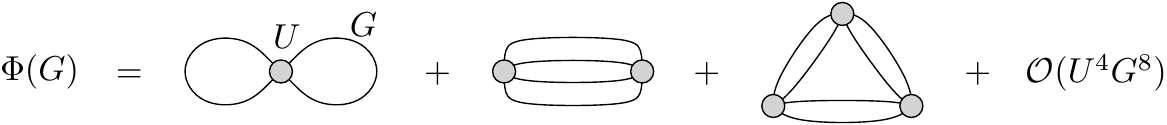}
  \end{center}
  \caption{\label{fig:Luttinger-Ward} Diagrammatic expansion of the
    Luttinger-Ward functional $\Phi[G]$.}
\end{figure}
This expansion is shown in Fig.~\ref{fig:Luttinger-Ward} and given by
\begin{eqnarray}
  \fl  \Phi[G] 
  = \frac{1}{8} \sum_{\gamma_1 \gamma_2} U_{\gamma_1 \gamma_2}
  G_{\gamma_1} G_{\gamma_2} \nonumber  \\
  - \frac{1}{48} \sum_{\gamma_1\gamma_2 \gamma_3 \gamma_4} U_{\alpha_1
    \alpha_2 \alpha_3 \alpha_4}
  U_{\alpha'_1 \alpha'_2 \alpha'_3 \alpha'_4} G_{\gamma_1} G_{\gamma_2}
  G_{\gamma_3} G_{\gamma_4}
  +  \mathcal{O}\left(U^3 G^6\right).
  \label{eq:Lutt-Ward-expand}
\end{eqnarray}
This was derived e.g. in Ref.~\cite{Wet07} for fermions and holds as
well for the anharmonic oscillator.  The individual addends are
proportional to $U^n G^{2n}$. In this expansion, it is apparent as
well that $\Phi$ does not depend on $C$. The higher derivatives
$\Gamma^{(n)}=\delta^n \Gamma/\delta G \dots \delta G$ and
$\Phi^{(n)}=\delta^n \Phi / \delta G \dots \delta G$ satisfy symmetry
relations that are analogous to Eq.~(\ref{eq:index-perm-symm}).

Physical quantities are obtained at vanishing external source
$J=0$. We mark those quantities by overlining, hence $\overline J=0,
\overline Z = Z[\overline J], \overline G=G[\overline J]$, etc. For
the Legendre transformed functions of the free variable $G$, the
physical state is defined equivalently by $\Gamma^{(1)}[\overline G] =
0$ and we write $\overline \Gamma = \Gamma[\overline G], \overline
\Phi{}^{(1)}=\Phi^{(1)}[\overline G]$ and so on. For systems with
spontaneously broken symmetry there may be several physical states.


\section{$C$-flow}
\label{sec:C-flow}


\subsection{Flow parameter, flow equations, initial conditions}
\label{sec:C-flow_flow-parameter}

In this section, we analyze 2PI fRG schemes where the flow
parameter is introduced in the free propagator, $C \rightarrow
C_\lambda$ (cf. Ref.~\cite{Dup05}). This method will be referred to as
``$C$-flow''. As in 1PI fRG schemes \cite{Met12}, we choose the
$\lambda$-dependence such that $C_{\lambda_\ix{i}}=0$ at the beginning
of the flow which ensures simple initial conditions for the flowing
moments of $\Phi$ (see below).  We enforce that at the end of the RG
flow $C_{\lambda_\ix{f}}=C$ such that the cutoff-free system of
interest is restored.

The $\lambda$-dependence of the free propagator $C_\lambda$ makes the
action and the functionals depend on $\lambda$: $S \rightarrow
S_\lambda, Z \rightarrow Z_\lambda$, etc. From now on, we suppress the
index $\lambda$ most times to avoid overloading the notation. The flow
of the functionals is given by
\begin{eqnarray}
  \dot Z[J] = \frac{\delta Z}{\delta J} \cdot \dot C^{-1},
  \\
  \dot W[J] = \frac{\dot Z}{Z} = W^{(1)}[J] \cdot \dot C^{-1},
  \\
  \dot \Gamma[G] = - \left. \dot W\right|_{J[G]} - \frac{\delta W}{\delta J}\cdot \dot J - \dot J \cdot G = - \left. \dot W \right|_{J[G]} = G \cdot \dot C^{-1}.  
\end{eqnarray}
Here, the dot denotes the derivative with respect to $\lambda$, like $\dot Z = d Z_\lambda / d \lambda$ and $\dot C^{-1} = d (C_\lambda^{-1}) / d \lambda$. We observe that $\dot \Gamma$ is independent of the interaction, $\dot \Gamma = \dot \Gamma_0$ (recall that $G$ is the free variable of $\Gamma$ and thus independent of the interaction). Consequently, the Luttinger-Ward functional $\Phi=\Gamma-\Gamma_0$ constitutes an invariant of the RG flow,
\begin{equation}
  \dot \Phi[G] = 0.
\end{equation}
This is expected since $\Phi$ does not depend on $C$ which carries the
flow parameter [cf. Eq.~(\ref{eq:Lutt-Ward-expand})]. The flow equations for the physical values of the
moments of $\Phi$ are thus
\begin{equation}
  \dot{\overline \Phi}{}^{(n)}_{\gamma_1 \dots \gamma_n} 
  = 
  \frac{d}{d \lambda} \Phi^{(n)}_{\gamma_1 \dots \gamma_n}[\overline G_\lambda]
  =
  \frac{1}{2} \sum_{\gamma} \overline \Phi{}^{(n+1)}_{\gamma_1 \dots
    \gamma_n \gamma} \dot {\overline G}_\gamma,
\end{equation}
or briefly $\dot{\overline\Phi}{}^{(n)}=\overline \Phi{}^{(n+1)} \cdot
  \dot{\overline G}$ for $n=0,1,2,\dots$, which is an infinite coupled hierarchy
comparable to those known from the 1PI fRG \cite{Met12}.

The propagator $\dot{\overline G}$ that enters the flow equations can be computed from $\overline G = (C^{-1}-\overline \Sigma)^{-1}$,
\begin{equation}
  \label{eq:G-bar-dot}
  \dot{\overline G} = - \overline G \left(\dot C^{-1} - \dot{\overline
    \Sigma} \right) \overline G,
\end{equation}
a result that depends in turn on $\dot{\overline \Sigma} = - \dot {\overline \Phi}{}^{(1)}$. This leads to a self-consistent flow equation for the self-energy,
\begin{equation}
  \dot {\overline{\Sigma}}
  = \overline \Phi{}^{(2)} \cdot \overline G \left(\dot C^{-1} -
    \dot{\overline \Sigma} \right) \overline G.
\end{equation}
An alternative approach to computing $\dot{\overline G}$ is given by $\dot{\overline G}= - \overline W{}^{(2)} \cdot \dot \Gamma^{(1)}|_{\overline G} = - \overline W{}^{(2)} \cdot \dot C^{-1}$ which is derived in Sect.~\ref{sec:U-flow-flow-param}, compare Eq.~(\ref{eq:G-bar-dot-alt}). This, however, requires the solution of the Bethe-Salpeter equation~(\ref{eq:Bethe-Salpeter}) for $W^{(2)}$.

We next discuss the initial conditions. From $\overline G=(C^{-1} -
\overline \Sigma)^{-1}$ and $C_{\lambda_\ix{i}}=0$ follows $\overline
G_{\lambda_\ix{i}}=0$. From Eq.~(\ref{eq:Lutt-Ward-expand}), we deduce
\begin{equation}
  \overline \Phi_{\lambda_\ix{i}} = 0,
  \qquad
  \overline \Phi{}^{(1)}_{\lambda_\ix{i}} = 0,
  \qquad
  \overline \Phi{}^{(2)}_{\lambda_\ix{i}} = U,
  \qquad
  \overline \Phi{}^{(3)}_{\lambda_\ix{i}} = 0,
\end{equation}
\begin{eqnarray}
  \overline{\Phi}{}^{(4),\lambda_\ix{i}}_{\gamma_1 \gamma_2 \gamma_3 \gamma_4} 
  &= 
  -\frac{1}{2} \left[\left(\left\{\left[\vphantom{U_{\alpha^\prime_1}}
          U_{\alpha_1 \alpha_2 \alpha_3 \alpha_4} 
          U_{\alpha'_1 \alpha'_2 \alpha'_3 \alpha'_4} 
          +\left. \zeta (\alpha_1\leftrightarrow \alpha_1^\prime)
          \right]\right.\right.\right.\right.
  \nonumber \\
  &\quad\quad\quad
  \left.\left.\left.\vphantom{U_{\alpha^\prime_1}} 
        +\zeta (\alpha_2 \leftrightarrow \alpha_2^\prime)\right\}
      +\zeta (\alpha_3\leftrightarrow \alpha_3^\prime)\right)
    +\zeta (\alpha_4\leftrightarrow \alpha_4^\prime)\right] ,
\end{eqnarray}
and more generally for $n=1,2,\dots$,
\begin{equation}
\label{eq:Dupuisfehler}
  \overline \Phi{}^{(2n-1)}_{\lambda_\ix{i}} = 0,
  \qquad
   0 \neq \overline \Phi{}^{(2n)}_{\lambda_\ix{i}} \sim U^n.
\end{equation}
Note that the nonvanishing result for $\overline
\Phi{}^{(4)}_{\lambda_\ix{i}}, \overline \Phi{}^{(6)}_{\lambda_\ix{i}},
\dots$ contradicts Ref.~\cite{Dup05} where it is claimed that
$\overline{\Phi}{}^{(n)}_{\lambda_\ix{i}} = 0$ for $n\ge 3$.


\subsection{Truncated flow equations and equivalence to self-consistent perturbation theory}
\label{sec:C_flow_truncated_flow_eq_equiv_to_scpt}

A straightforward truncation of the infinite hierarchy of flow
equations for the $\overline \Phi{}^{(n)}$ leading to a closed set of equations
is obtained by neglecting the flow of $\overline \Phi{}^{(m)}$ (and
higher moments) for some fixed given $m$ by setting $\overline
\Phi{}^{(m)}_\lambda=\overline \Phi{}^{(m)}_{\lambda_\ix{i}}$.  As
$\overline \Phi{}^{(m)}_{\lambda_\ix{i}}$ vanishes for odd $m$,
truncations at level $m=2l$ and $m=2l+1$ are equivalent, yielding both
$\overline \Phi{}^{(2l)}_\lambda = \overline
\Phi{}^{(2l)}_{\lambda_\ix{i}} \sim U^l$.  As in 1PI fRG, this way of
truncating is motivated from a weak coupling perspective: if $U$ is
sufficiently small, the impact of the higher moments of
$\overline{\Phi}$ on the flow of the lower ones can be neglected.  The
relation of this truncation to established self-consistent
perturbative approximation schemes is discussed next.

The choices $m=0$ and $1$ are pointless since they lead to $\overline \Phi=0, \overline \Phi{}^{(1)} = -\overline \Sigma = 0$ during all of the flow. Let us hence consider the case $m=2$ or $3$, which means setting $\overline \Phi{}^{(2)}_\lambda=U$.  The resulting flow equation $\dot {\overline \Phi}{}^{(1)} = -\dot{\overline \Sigma} = U\cdot \dot{\overline G}$ can directly be integrated leading to $\overline \Sigma = -U\cdot \overline G$ which is precisely the Hartree-Fock self-consistency equation for the self-energy. Integration of the flow equation $\dot{\overline \Phi} = -\overline \Sigma \cdot \dot{\overline G} = \overline G \cdot U \cdot \dot{\overline G}$ in turn yields $\overline \Phi = \frac{1}{2} \overline G \cdot U \cdot \overline G$, which is the first order perturbation theory result for the Luttinger-Ward functional, compare Eq.~(\ref{eq:Lutt-Ward-expand}).

Diagrammatic approximations to the Luttinger-Ward functional combined with the self-consistency equation $\Sigma=-\Phi^{(1)}[G]$ are well known \cite{Bay62}. They lead to conserving approximations \cite{Bay62,Bay61}. In particular, the first order approximation to $\Phi$ reproduces self-consistent Hartree-Fock, that is mean-field theory. More generally, the $n$th order approximation to $\Phi$ leads to $n$th order self-consistent perturbation theory for the self-energy. This means that the self-energy is computed from diagrams up to order $U^n$, with lines denoting full propagators $G$ that are determined self-consistently.

Now, we observe in general that truncating our set of flow equations
at level $m=2n$ (or, equivalently, $m=2n+1$) and solving the remaining
flow equations results precisely in $n$th order self-consistent
perturbation theory. This holds independently of which
$\lambda$-dependence is chosen for $C$, as long as
$C_{\lambda_\ix{i}}=0$ and $C_{\lambda_\ix{f}}=C$.  In particular,
this is independent of whether or not $C_{\lambda}$ regularizes
diagrammatic expressions which are infrared divergent with bare
propagators. Furthermore, this is true not only at $\lambda_\ix{f}$ but
even during all of the flow (if perturbation theory is formulated with
$C_\lambda$ instead of $C$).  For the proof let us denote the $n$th
order perturbation theory approximation to $\Phi[G]$ by $A[G]$.  From
$A$ being independent of $C$ follows $\dot A[G] = 0$ and
$\dot{\overline A}{}^{(l)}=\overline A{}^{(l+1)} \cdot \dot {\overline
  G}$. The initial conditions are
\begin{equation}
  \overline A{}^{(l)}_{\lambda_\ix{i}} =   \overline
  \Phi{}^{(l)}_{\lambda_\ix{i}}, l=0,1,\dots,2n,
  \qquad
  \overline A{}^{(l)}_{\lambda_\ix{i}} = 0, l\ge 2n+1.
\end{equation}
Obviously, the hierarchy of flow equations satisfied by the $\overline
A{}^{(l)}$ is finite and identical to the truncated exact hierarchy. As
$\overline{A}$ and the truncated $\overline{\Phi}$ fulfill the same
differential equations and initial conditions, they are the same. This
concludes the proof.

We thus derived a fundamental equivalence of two different approaches
to the many-body problem: the solution of the 2PI $C$-flow equations
truncated at level $2n$ gives exactly self-consistent perturbation
theory to level $n$.  It is rather remarkable that the resummation
inherent to the RG procedure leads in the most natural truncation of
the 2PI $C$-flow to a well known conserving approximation.  This must
be contrasted to most other truncated fRG schemes. E.g. in 1PI fRG, the
lowest order truncation for the self-energy is not equivalent to
self-consistent Hartree-Fock \cite{Met12}. Also, the other 2PI fRG schemes discussed here (internally closed $C$-flow and 
two $U$-flow schemes, see below) are not equivalent to self-consistent perturbation theory.


\subsection{Internally closed set of flow equations}

Given the equivalence of truncated 2PI $C$-flow and self-consistent
perturbation theory, we face the question whether other approximations
to the $C$-flow are conceivable which lead beyond this known
approximation to the quantum many-body problem.  Such an approximation
was studied in Ref.~\cite{Dup05}, however not motivated by going
beyond the above derived equivalence.\footnote{This equivalence is not
  observed in Ref.~\cite{Dup05}, presumably due to the wrong initial
  conditions [see the sentence following our
  Eq.~(\ref{eq:Dupuisfehler})].}  The idea is to truncate the
hierarchy of flow equations for the $\overline{\Phi}{}^{(n)}$ by finding
an approximate expression for $\overline{\Phi}{}^{(3)}$ in terms of
$\overline{\Phi}{}^{(2)}$. The procedure is based on the leading (second
order) perturbative contribution to $\overline \Phi{}^{(3)}$,
\begin{eqnarray}
  \left. \overline{\Phi}{}^{(3)}_{\gamma_1 \gamma_2 \gamma_3} \right|_\ix{2nd}
  &=
  -\frac{1}{2} \sum_{\gamma_4}  
  \left(\left\{\left[\vphantom{U_{\alpha^\prime_1}}\right.\right.\right. 
  U_{\alpha_1 \alpha_2 \alpha_3 \alpha_4} U_{\alpha'_1 \alpha'_2
    \alpha'_3 \alpha'_4}
  +\left. \zeta (\alpha_1\leftrightarrow \alpha'_1) \right] \nonumber
  \\  
  &\hspace{6em}
  \left.\left.\vphantom{U_{\alpha^\prime_1}} 
      + \zeta (\alpha_2 \leftrightarrow \alpha'_2)\right\}
    + \zeta (\alpha_3 \leftrightarrow \alpha'_3)\right)
  \overline{G}_{\gamma_4},
  \label{eq:Phi3_2nd}
\end{eqnarray}
compare to Eq.~(\ref{eq:Lutt-Ward-expand}). In Ref.~\cite{Dup05}, it
is proposed to replace $U \to \overline{\Phi}{}^{(2)}$ in the expression
for $\left. \overline{\Phi}{}^{(3)} \right|_\ix{2nd}$. We observe that
this idea contains some ambiguity: while $U$ satisfies the full index
permutation (anti-)symmetry $U_{\alpha_1 \alpha_2 \alpha_3 \alpha_4} =
\zeta^P U_{\alpha_{P1} \alpha_{P2} \alpha_{P3} \alpha_{P4}}$, the
symmetry of $\overline{\Phi}{}^{(2)}$ is only analogous to
Eq.~(\ref{eq:index-perm-symm}). In Ref.~\cite{Dup05}, some specific
order of indices of the interaction amplitudes $U$ in
Eq.~(\ref{eq:Phi3_2nd}) is chosen before replacing $U \to
\overline{\Phi}{}^{(2)}$. Other choices would lead to other
approximations. We consider it appropriate to replace instead
\begin{equation}
\fl U_{\alpha_1 \alpha_2 \alpha_3 \alpha_4}\to
\overline{\Phi}{}^{(2),\ix{sym}}_{\alpha_1 \alpha_2
  \alpha_3 \alpha_4} =\frac{1}{3}\left(
  \overline{\Phi}{}^{(2)}_{\alpha_1 \alpha_2 \alpha_3 \alpha_4}+
  \overline{\Phi}{}^{(2)}_{\alpha_2 \alpha_3 \alpha_1 \alpha_4} +
  \overline{\Phi}{}^{(2)}_{\alpha_3 \alpha_1 \alpha_2 \alpha_4} \right),
\end{equation}
where $\overline{\Phi}{}^{(2),\ix{sym}}$ has the full index permutation (anti-)symmetry. The resulting flow equation,
\begin{eqnarray}
  \dot{\overline{\Phi}}{}^{(2)}
  = \left(
    \left.\overline{\Phi}{}^{(3)} \right|_{\ix{2nd}}
  \right)_{U \rightarrow \overline{\Phi}{}^{(2),\ix{sym}}} \cdot 
  \dot{\overline{G}},
\end{eqnarray}
closes the hierarchy. Note that the (anti-)symmetrization of $\overline \Phi{}^{(2)}$ 
on the right-hand-side of the flow equation is indeed required since $\overline \Phi{}^{(2)}$ 
resulting from the flow equation does not have the full (anti-)symmetry. We refer to this scheme as ``internally closed $C$-flow''.

Due to $\overline \Phi{}^{(2)}_{\lambda_\ix{i}} = U$ the flow of $\left(
\left.\overline{\Phi}{}^{(3)} \right|_{\ix{2nd}} \right)_{U \rightarrow
  \overline{\Phi}{}^{(2),\ix{sym}}}$ starts at
$\left.\overline{\Phi}{}^{(3)} \right|_{\ix{2nd}}$. Integrating the flow
equations then leads to
\begin{eqnarray}
  \overline{\Phi}{}^{(2)}_{\lambda_\ix{f}} &= \left.\overline{\Phi}{}^{(2)}
  \right|_{\ix{2nd}} + \mathcal{O}(U^3 \overline G^4),
  \\
  \overline \Sigma_{\lambda_\ix{f}} &= \left.\overline{\Sigma}
  \right|_{\ix{2nd}} + \mathcal{O}(U^3 \overline G^5),
  \\
  \overline \Phi_{\lambda_\ix{f}} &= \left.\overline{\Phi}
  \right|_{\ix{2nd}} + \mathcal{O}(U^3 \overline G^6),
\end{eqnarray}
where $\left.\overline{\Phi} \right|_{\ix{2nd}}$ denotes the
contributions to $\overline \Phi$ in first and second order in $U$,
compare Eq.~(\ref{eq:Lutt-Ward-expand}).  Consequently, the solution
of the internally closed $C$-flow comprises second order
self-consistent perturbation theory. For the self-energy this means
explicitly
\begin{equation}
  \overline \Sigma_{\lambda_\ix{f}} = \overline \Sigma_\ix{exact} +
  \mathcal{O}(U^3 \overline G^5_\ix{2SCPT}),
\end{equation}
where $\overline G_\ix{2SCPT}$ denotes the full propagator of the
physical state in second order self-consistent perturbation theory.


\subsection{Computing the thermodynamic potential}
\label{sec:compOm}

One of the key observables depending on the interaction is the
expectation value of the energy, which follows from the thermodynamic
potential $\Omega[G]=\Gamma[G]/\beta$ with physical value
\begin{equation}
  \overline \Omega = -\frac{1}{\beta} \ln \overline Z.
\end{equation}
For instance, the ground state energy is $E_\ix{gs}=\lim_{\beta \to \infty} \overline{\Omega}$.

The thermodynamic potential can be computed from the values of
$\overline \Phi$ and $\overline \Sigma$ that result at the end of the
truncated or internally closed $C$-flow (or which follow in
perturbation theory to $\Phi[G]$ from the self-consistency equation
$\Phi^{(1)}[\overline G]=-\overline \Sigma = \overline G^{-1} -
C^{-1}$).  The physical value of the effective action is
\begin{equation}
  \overline \Gamma 
  = \overline \Phi +  \Gamma_0[\overline G] 
  = \overline \Phi - \frac{\zeta}{2} \big[ \tr \ln (-\overline G)
  -\tr(C^{-1}\overline G-1) \big].
\end{equation}
With $\frac{\zeta}{2}\tr(C^{-1} \overline G-1)=(C^{-1}-\overline G^{-1}) \cdot \overline G = \overline\Sigma \cdot \overline G$, this means
\begin{equation}
  \label{eq:Omega-pert}
  \overline \Omega = \frac{1}{\beta} \left[- \frac{\zeta}{2} \tr \ln
    (-\overline G) 
    + \overline \Phi + \overline \Sigma \cdot \overline G \right].
\end{equation}

For example in $C$-flow truncated at level $m=2$, that is in lowest
order perturbation theory to $\Phi[G]$, the resulting self-consistency
equation $\overline \Sigma = - U \cdot \overline G$ corresponds to the
mean-field approximation, $\overline G = \overline G_\ix{MF}$. From
$\Phi_\ix{MF}[\overline G_\ix{MF}] + \overline \Sigma_\ix{MF}\cdot \overline
G_\ix{MF} = - \frac{1}{2} \overline G_\ix{MF} \cdot U \cdot \overline
G_\ix{MF}$ follows
\begin{eqnarray}
  \overline \Omega_\ix{MF} 
  &= \frac{1}{\beta} \left[- \frac{\zeta}{2} \tr \ln (- \overline G_\ix{MF})
    - \frac{1}{2} \overline G_\ix{MF} \cdot U \cdot \overline
    G_\ix{MF} \right] 
  \\
 \label{eq:Omega-MF}
 &= \Omega_0[C]_{C \rightarrow \overline G_\ix{MF}} -
 \frac{1}{2\beta} \overline G_\ix{MF} \cdot U \cdot \overline G_\ix{MF},
\end{eqnarray}
which is the well known mean-field result.  In order to evaluate
$\Omega_0[C]_{C \rightarrow \overline G_\ix{MF}}$ one replaces the
bare single-particle energies by mean-field dressed energies in the
analytic result for $\Omega_0[C]=-(\ln \overline Z_0)/\beta$. For the anharmonic
oscillator, e.g., one replaces $\omega_\ix{G} \rightarrow
\sqrt{\omega_\ix{G}^2 + \overline \Sigma_\ix{MF}}$ in $\ln \overline
Z_0 = - \ln [2 \sinh(\beta \omega_\ix{G}/2) ]$.

Higher order truncations of the flow produce a frequency dependent $\overline G$ which makes the evaluation of $\tr \ln (- \overline G)$ less trivial. Then, it is convenient to compute the thermodynamic potential from a flow equation instead. Consider any approximation to the flow equations for $\overline \Phi{}^{(1)},\overline \Phi{}^{(2)},\dots$ that result in some function $\overline \Phi{}^{(1)}_\lambda=-\overline \Sigma_\lambda$. This function allows to compute the flow of $\overline{G}=\left(C^{-1}-\overline{\Sigma}\right)^{-1}$ and of $\overline{\Phi}$ as
\begin{equation}
  \dot{\overline{G}}
  =-\overline{G}\left(\dot
    C^{-1}-\dot{\overline{\Sigma}}\right)\overline{G},
  \qquad
  \dot{\overline{\Phi}} =-\overline{\Sigma}\cdot\dot{\overline{G}}.   
\end{equation}
In the same approximation, the flow of $\overline{\Omega}$ is
\begin{eqnarray}
  \dot{\overline{\Omega}}
  =\frac{1}{\beta}\left( \dot{\overline{\Phi}}+\frac{d}{d\lambda}
    \Gamma_0[\overline{G}]\right) 
  =\frac{1}{\beta}
  \left(-\overline{\Sigma}\cdot\dot{\overline{G}}
    +\dot{\Gamma}_0[\overline{G}] 
    + \Gamma_0^{(1)}[\overline{G}]\cdot\dot{\overline{G}}\right).
\end{eqnarray}
Now $\dot{\Gamma}_0[\overline{G}]=\overline G \cdot \dot C^{-1}$ and $\Gamma_0^{(1)}[\overline{G}]=C^{-1}-\overline G^{-1} = \overline \Sigma$ result in
\begin{equation}
  \dot{\overline{\Omega}} =\frac{1}{\beta} \overline G \cdot \dot
  C^{-1}. 
\end{equation}
We observe that the form of this flow equation is independent of the approximation chosen for the flow of $\overline{\Sigma}_\lambda$. Hence, it is valid as well in case of the exact flow, compare Ref.~\cite{Dup05}.

The initial condition
\begin{equation}
  \overline \Omega_{\lambda_\ix{i}} 
  = \frac{1}{\beta} \left( \overline \Phi
    + \Gamma_0[\overline G]
  \right)_{\lambda_\ix{i}}
  =
  -\frac{\zeta}{2\beta}\big[\tr \ln(-\overline G) -\tr(C^{-1} \overline G
  -1) \big]_{\lambda_\ix{i}}   
\end{equation}
is difficult to handle since $\overline G_{\lambda_\ix{i}}=0$ leads to a divergence of $\ln(-\overline G)$. Following Ref.~\cite{Dup05}, we study instead
\begin{equation}
  \Delta \overline \Omega
  = \overline \Omega - \Omega_0[C]
  = \overline \Omega + \frac{\zeta}{2\beta} \tr\ln (-C)
\end{equation}
with initial condition $\Delta \overline \Omega_{\lambda_\ix{i}}=0$ and flow equation
\begin{equation}
  \label{eq:DeltaOmega-flow}
  \Delta \dot{\overline \Omega}
  =
  \frac{d}{d\lambda} \Delta \overline \Omega 
  =   \dot{\overline \Omega}
  + \frac{\zeta}{2\beta} \frac{d}{d\lambda} \tr \ln (-C)
  =\frac{1}{\beta} (\overline G -  C)\cdot \dot C^{-1}.
\end{equation}
At the end of the flow, we recover the thermodynamic potential as $\overline \Omega_{\lambda_\ix{f}} = \Omega_0[C] \big|_{\lambda_\ix{f}} +\Delta \overline \Omega_{\lambda_\ix{f}}.$  


\subsection{Application to the anharmonic oscillator}

We test the performance of $C$-flow truncated at level $m=2,4$ or internally closed on the quantum anharmonic oscillator (with $\zeta=+$ and $\alpha=\tau$). Due to translational invariance in imaginary time, it is preferable to work in frequency representation. Hence, we Fourier transform the $m$-point functions to bosonic Matsubara frequencies $\omega_n=\frac{2\pi}{\beta} n$,
\begin{eqnarray}
  \label{eq:FT_1}
  \fl
  W^{(m)}_{n_1 n'_1 \dots n_m n'_m}
  &= \int_{\tau_1} \!\dots\! \int_{\tau'_m}
  e^{-i \omega_{n_1} \tau_1 \dots - i \omega_{n_m} \tau_m}
  W^{(m)}\left(\tau_1,\dots,\tau'_m\right)
  e^{-i \omega_{n'_1} \tau'_1 \dots - i \omega_{n'_m} \tau'_m},
  \\ \label{eq:FT_2}
  \fl
  \Phi^{(m)}_{n_1 n'_1 \dots n_m n'_m}
  &= \int_{\tau_1} \!\dots\! \int_{\tau'_m}
  e^{+i \omega_{n_1} \tau_1 \dots + i \omega_{n_m} \tau_m}
  \Phi^{(m)}\left(\tau_1,\dots,\tau'_m\right)
  e^{+i \omega_{n'_1} \tau'_1 \dots + i \omega_{n'_m} \tau'_m},
\end{eqnarray}
where $\int_\tau=\int_0^\beta d\tau$. Mind the choice of signs in the exponents. Here, $W^{(m)}$ is a representative of propagators (including $G$), whereas $\Phi^{(m)}$ represents vertices (including $\Sigma$). For the free inverse propagator one obtains
\begin{equation}
 \left(C^{-1}\right)_{n n'}
 =-\beta \delta_{n+n',0} \left(i \omega_n i \omega_{n^\prime}+\omega_\ix{G}^2\right)
 =-\beta \delta_{n+n',0} \left(\omega_n^2+\omega_\ix{G}^2\right) .
\end{equation}
More generally, all relevant functions are proportional to $\beta \delta_{n_1+\dots,0}$. We refer to the remaining part by skipping the last index, e.g.
\begin{equation}
  \Sigma_{n n^\prime}
  =\beta \delta_{n+n^\prime,0} \Sigma_n,
  \qquad
  \Phi^{(2)}_{n_1 n'_1 n_2 n'_2}
  =\beta \delta_{n_1+n'_1+n_2+n'_2,0} \Phi^{(2)}_{n_1
    n'_1 n_2} .
\end{equation}

For the flow parameter dependence of $C$, we choose specifically
\begin{equation}
  C^\lambda_n = - \left(\omega_n^2 +\omega_\ix{G}^2 + \lambda\right)^{-1},
\end{equation}
with $\lambda_\ix{i}=\infty$ and $\lambda_\ix{f}=0$. This can be
considered as an analogue to the so-called hybridization flow
parameter for single-level quantum impurity problems
\cite{Jak10a,Jak10b,Kar10}, which has the form $C^\lambda_n = \left[i
  \omega_n -\varepsilon + i\sgn(\omega_n) (\Gamma + \lambda)
\right]^{-1}$ for a system in grand canonical equilibrium.  Here,
$\varepsilon$ denotes the on-site energy of the level and $\Gamma$ the
hybridization to the leads.  Generally speaking, the results following
from the solution of truncated fRG flow equations depend on the cutoff
choice (in contrast to the solution of the infinite hierarchy of
equations) \cite{Met12}.  Therefore, it is an interesting observation
that the results of truncated $C$-flow are completely independent of
the particular choice of the flow parameter as they are equivalent to
self-consistent perturbation theory.  This does not hold for the
internally closed $C$-flow.  By comparison to exact results, it was
shown for selected fermionic many-body models that the hybridization
flow parameter is a good choice \cite{Jak10b,Kar10} in 1PI fRG. This
provides the motivation to use its analogue for the problem at hand. We
here do not further investigate the influence of the cutoff choice on
the quality of the results.

Dyson's equation yields
\begin{equation}
  \overline{G}_n 
  = \left[(C^{-1})_n - \overline{\Sigma}_n\right]^{-1}
  = -\left( \omega_n^2 + \omega_\ix{G}^2 + \lambda +
    \overline{\Sigma}_n \right)^{-1} .
\end{equation}
Due to $\dot C^{-1}_n= -1$  the flow equation~(\ref{eq:DeltaOmega-flow}) for the thermodynamic potential takes the form
\begin{equation}
  \label{eq:AnhOsc_Delta-Omega-flow}
  \Delta \dot{\overline{\Omega}}
  = -\frac{1}{2\beta} \sum_{n} \left(\overline{G}_{n}-C_{n}\right).
\end{equation}
The flow of the self-energy $\overline \Sigma = - \overline \Phi{}^{(1)}$ reads
\begin{equation}
  \label{eq:AnhOsc_Sigma-flow}
  \dot{\overline{\Sigma}}_{n_1} 
  =
   -\frac{1}{2\beta} \sum_{n_3} \overline{\Phi}{}^{(2)}_{n_1,-n_1,n_3} 
   \dot{\overline{G}}_{n_3},
\end{equation}
with
\begin{equation}
  \dot{\overline{G}}_n 
  = - \overline{G}_n \left[(\dot{C}^{-1})_{-n} - \dot{\overline{\Sigma}}_{-n}\right] \overline{G}_n .
\end{equation}
The initial conditions are $\overline{\Sigma}{}_n^{\lambda_\ix{i}}=0$ and $\overline{\Phi}{}^{(2),\lambda_\ix{i}}_{n_1 n'_1 n_2}=g$. In case of $C$-flow truncated at level $m=4$, the solution is self-consistent second order perturbation theory,
\begin{equation}
  \label{eq:Sigma_Mats_C_D3O}
  \overline{\Sigma}_{n_1}
  = -\frac{g}{2\beta} \sum_{n_2} \overline{G}_{n_2} 
  + \frac{g^2}{6 \beta^2} \sum_{n_2 n_3} \overline{G}_{n_2}
  \overline{G}_{n_3}  \overline{G}_{-n_1 -n_2 -n_3}.
\end{equation}
This equation is solved self-consistently for each $\lambda$ (which is numerically faster and more accurate than integrating the flow of $\overline \Sigma$, $\overline \Phi{}^{(2)}$ and $\overline \Phi{}^{(3)}$) and the result is used to calculate the flow of $\Delta \overline{\Omega}$.

For the internally closed $C$-flow, Eq.~(\ref{eq:AnhOsc_Sigma-flow}) is solved numerically together with
\begin{equation}
  \label{eq:Phi2_flow_eq_IC2O}
  \fl \dot{\overline{\Phi}}{}^{(2)}_{n_1 n'_1 n_2}\!\!
  = \!-\frac{1}{\beta} \!\sum_{n_3}\!\! \left.\left\{
      \overline{\Phi}{}^{(2),\ix{sym}}_{n_1,n_2,n_3}
      \overline{\Phi}{}^{(2),\ix{sym}}_{n'_1,n'_2,n'_3} 
      \overline{G}_{n_1+n_2+n_3} 
      + ( n_1 \leftrightarrow n'_1)
    \right\}\right|^{n'_2=-n_1-n'_1-n_2}_{n'_3=-n_3}
  \dot{\overline{G}}_{n_3},
\end{equation}
\begin{equation}
  \overline{\Phi}{}^{(2),\ix{sym}}_{n_1,n_2,n_3}
  = \frac{1}{3} \left( \overline{\Phi}{}^{(2)}_{n_1,n_2,n_3} 
    + \overline{\Phi}{}^{(2)}_{n_2,n_3,n_1} 
    + \overline{\Phi}{}^{(2)}_{n_3,n_1,n_2} \right) .
\end{equation}
Eq.~(\ref{eq:Phi2_flow_eq_IC2O}) exploits the symmetry $\overline{\Sigma}_{n_1}=\overline{\Sigma}_{-n_1}$ which implies $\overline{G}_{n_1}=\overline{G}_{-n_1}$. It follows from Eq.~(\ref{eq:index-perm-symm}), which translates to $\overline{\Phi}{}^{(2)}_{n_1 n'_1 n_2}=\overline{\Phi}{}^{(2)}_{n'_1 n_1 n_2}$ in frequency space, via Eq.~(\ref{eq:AnhOsc_Sigma-flow}).

We restrict our numerical investigations to the case of vanishing
temperature and thus take the limit $\beta \to \infty$. The Matsubara
frequencies become continuous then, $\omega_n \to \nu$, and sums turn
into integrals, $\frac{1}{\beta} \sum_n \to \int \frac{d\nu}{2\pi}$.
Transforming, for instance, Eq.~(\ref{eq:Sigma_Mats_C_D3O}), one finds
explicitly
\begin{equation}
\label{eq:Sigma_Mats_C_D3O_cont}
  \fl \overline{\Sigma}(\nu) 
  = -\frac{g}{2}\int_{-\infty}^\infty \frac{d\nu_1}{2\pi}
  \overline{G}(\nu_1)
  + \frac{g^2}{6} \int_{-\infty}^\infty\!\! \frac{d\nu_2}{2\pi} 
  \int_{-\infty}^\infty\!\! \frac{d\nu_3}{2\pi} \,
  \overline{G}(\nu_2) \overline{G}(\nu_3)\overline{G}(-\nu\!-\!\nu_2\!-\!\nu_3).
\end{equation}

We here focus on two observables.  One is the shift in the ground
state energy induced by the anharmonicity, $e_\ix{gs}= E_\ix{gs} -
E^0_\ix{gs} = \lim_{\beta \to \infty} \Delta \overline{\Omega}$, with
the non-interacting ($g=0$) ground state energy $ E^0_\ix{gs}$.  The
other is the fluctuation of the position $\varrho=\left \langle x^2
\right \rangle$ which corresponds to the particle density in a
fermionic many-body problem.  In the following we therefore refer to
$\varrho$ as the density.  At the end of the flow, it can be extracted
from the self-energy as
\begin{equation}
  \fl
  \varrho
  =-\overline{G}(0,0)
  = \int_{-\infty}^{\infty} \frac{d\nu}{2\pi} 
  \frac{1}{\nu^2 + \omega_\ix{G}^2 + \overline{\Sigma}(\nu)}
  \approx
  \int_0^{\nu_{m_\ix{len}}} \frac{d\nu}{\pi}
  \frac{1}{\nu^2+\omega_\ix{G}^2+\overline{\Sigma}(\nu)}
  + \frac{1}{\pi\nu_{m_\ix{len}}}.
\end{equation}
Here, the last integral refers to the numerical evaluation on a grid of frequencies $\nu_n$ with $n=-m_\ix{len}, \dots, m_\ix{len}$ and $\nu_n=-\nu_{-n}$ (details on the grid are given in \ref{App:Num_implementation}). The integral is then calculated by an interpolator. The very last addend is an approximation to the missing integral part $-\int_{\nu_{m_\ix{len}}}^\infty \!\!\frac{d\nu}{\pi} \overline{G}(\nu)$.

In case of $C$-flow truncated at level $m=2$, the solution is
mean-field theory. At zero temperature, the self-consistency equation
$\overline \Sigma_\ix{MF} = - U \cdot \overline G_\ix{MF}$ takes the
form
\begin{equation}
  \overline \Sigma_\ix{MF} = \frac{g}{2} \langle x^2 \rangle_\ix{MF} =
  \frac{g}{4} \frac{1}{\sqrt{\omega_\ix{G}^2 + \overline \Sigma_\ix{MF}}}.
\end{equation}
In this case, we do not integrate numerically any flow but solve the
self-consistency equation directly at $\lambda_\ix{f}$ and compute the
thermodynamic potential from Eq.~(\ref{eq:Omega-MF}) (taking $\beta
\to \infty$).  This leads to the regular mean-field result
\cite{Hed04}
\begin{equation}
  e^\ix{MF}_\ix{gs} = \frac{1}{2} \sqrt{\omega_\ix{G}^2 + \overline
    \Sigma_\ix{MF}} - \frac{1}{2g}\overline
  \Sigma_\ix{MF}^2 - \frac{\omega_\ix{G}}{2}.
\end{equation}

For the internally closed $C$-flow, $\overline{\Phi}{}^{(2)}_\lambda$
depends on three frequencies. Consequently, the number of flow
equations scales $\sim m_\ix{len}^3$. As one sum must be performed on
the right-hand-side of Eq.~(\ref{eq:Phi2_flow_eq_IC2O}), the effort of
computing all the right-hand-sides of the set of differential
equations in each step of the flow scales $\sim m_\ix{len}^4$. This
severely limits the accessible $m_\ix{len}$ in this scheme. In
comparison, the regular $C$-flow schemes truncated at arbitrary order
$m$ always produce $\sim m_\ix{len}$ number of flow equations. This is
because only $\Delta \overline{\Omega}$ and $\overline{\Sigma}(\nu)$
flow in the regular schemes as the flow of $\overline{\Sigma}(\nu)$
can be determined self-consistently. Nevertheless, there is (of
course) an increasing cost for calculating increasing orders $m$: the
effort to calculate the right-hand-side of the self-consistency
equation for $\overline{\Sigma}(\nu)$ increases. For example at $m=4$,
the effort to calculate all equations for one step of the flow is
$\sim m_\ix{len}^2$---this can be managed by pre-calculating one of
the $d\nu_2,d\nu_3$-integrals in Eq.~(\ref{eq:Sigma_Mats_C_D3O_cont})
before actually calculating the right-hand-side of the equation
itself.


\subsection{Numerical results for the anharmonic oscillator}

\begin{figure}
 \includegraphics[width=0.5\textwidth]{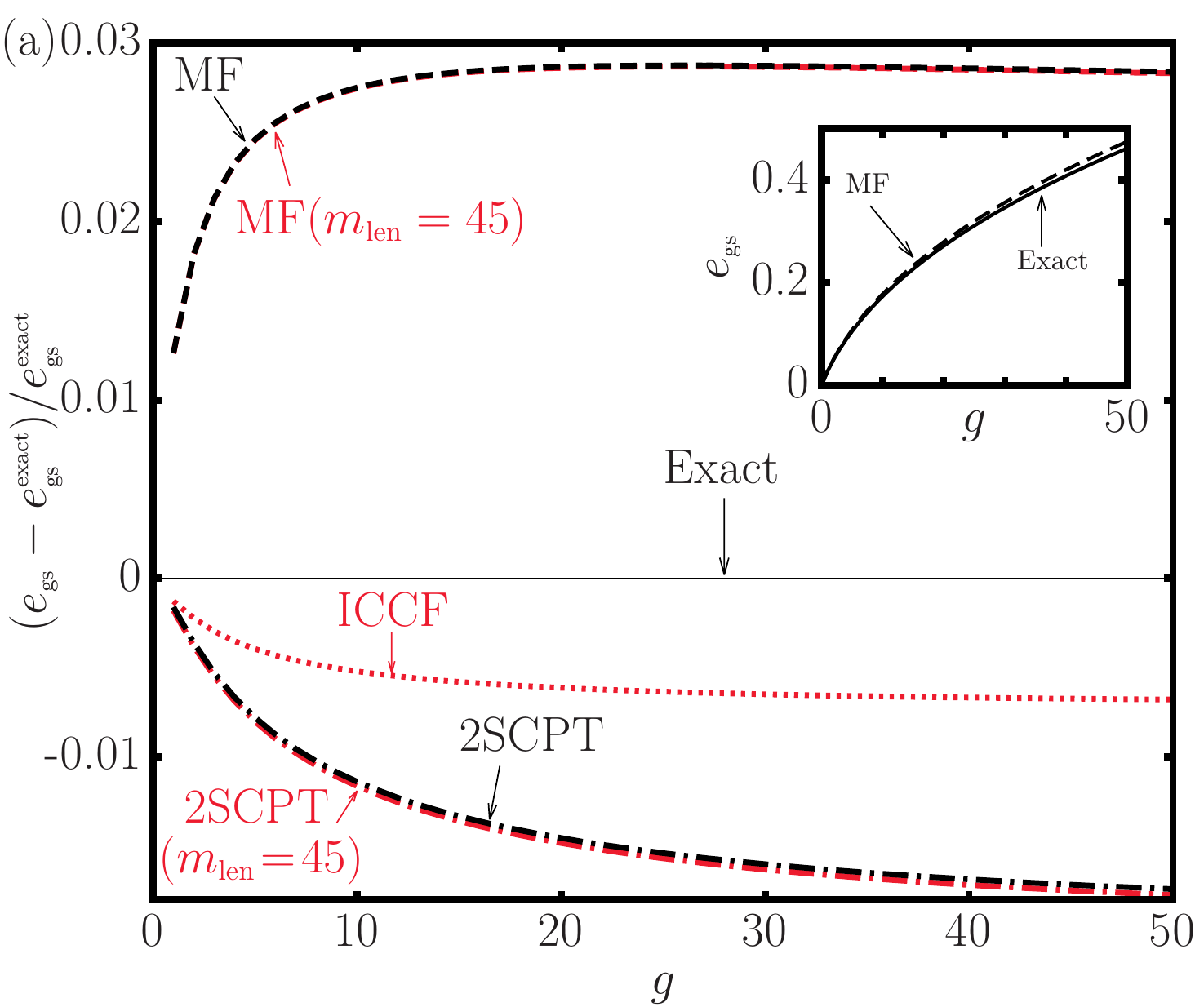}
 \includegraphics[width=0.5\textwidth]{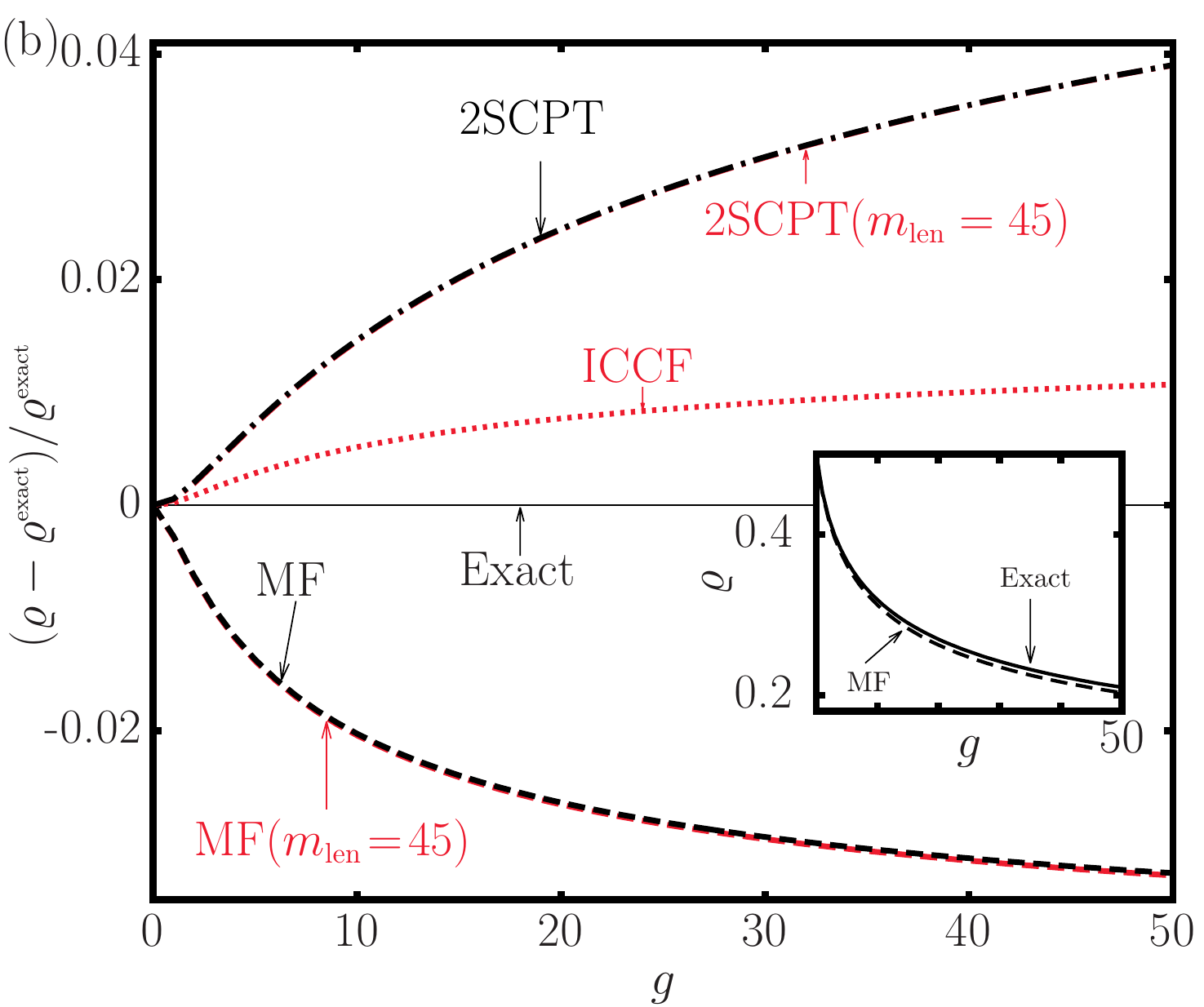}
  \caption{\label{fig:Paper_QAOsc_C-flow} Numerical results for the
    ground state energy $e_\ix{gs}$ (a) and the density $\varrho$ (b)
    as a function of the strength of the anharmonicity $g$.  The lines
    show data obtained from $C$-flow truncated at level $m=2$
    (dashed), that is mean-field theory (MF), from $C$-flow truncated
    at level $m=4$ (dashed-dotted), that is second order
    self-consistent perturbation theory (2SCPT), and from internally
    closed $C$-flow (dotted; ICCF).  The main plots show the relative
    difference to the exact result, whereas the insets show the
    absolute mean-field and exact curves.}
\end{figure}

For all numerical considerations, $\omega_\ix{G}$ is set to $1$ in this paper. Fig.~\ref{fig:Paper_QAOsc_C-flow} shows numerical results for the ground state energy shift and the density as functions of the coupling constant $g$. The data obtained from $C$-flow truncated at levels $m=2,4$ or internally closed are compared to the exact results. As the actual (``absolute'') curves are very close to each other, the main plots show the relative differences to the exact result.

We compute the numerically exact data by matrix diagonalization: 
the 
full Hamiltonian (with quartic anharmonicity) can be expressed as a matrix in the basis of the energy eigenstates $\left| E_n^0 \right>$ of the unperturbed harmonic oscillator. Diagonalizing this matrix yields the energy eigenvalues 
and eigenstates of the anharmonic system.
The ground state density is then obtained as $\varrho_\ix{gs}^\ix{exact}=\left< E_0 \right| \hat{x}^2 \left| E_0 \right>$, where the matrix $(\hat{x}^2)_{m,n}=\left<E_m^0\right| \hat{x}^2 \left| E_n^0\right>$ in the unperturbed eigenbasis only has the following non-zero elements: $(\hat{x}^2)_{n,n}=n+\frac{1}{2}$ and $(\hat{x}^2)_{n+2,n}=(\hat{x}^2)_{n,n+2}=\frac{1}{2} \sqrt{n+1}\sqrt{n+2}$. This can easily be proven from the ladder operator representation $\hat{x}=\left(\hat{a}+\hat{a}^\dagger\right)/\sqrt{2}$. A matrix size of $200 \times 200$ is found to be sufficient for converged numerics.

In all numerical $C$-flow computations, the frequency grid covers the
range from $\nu_{-m_\ix{len}} = -5120$ to $\nu_{m_\ix{len}}=5120$ with
$2m_\ix{len}+1$ grid points, see \ref{App:Num_implementation} for
details on the grid.  The truncated $C$-flow is computed with
$m_\ix{len}= 500$, while the numerically much more demanding
internally closed $C$-flow is solved with $m_\ix{len}=45$. In the
latter case, comparison to $m_\ix{len}=35$ and $40$ data suggests that
convergence was reached on the scale of the figure.\footnote{If at
  all, the energy curve is expected to move a little closer towards
  the exact data for higher $m_\ix{len}$.} This is also supported by
the observation that data from truncated $C$-flow computed at
$m_\ix{len}=45$ are close to their converged counter-parts. For all
$C$-flow methods, the flow starts at
$\lambda_\ix{i}^\ix{num}=10^{16}$. Convergence with respect to
$\lambda_\ix{i}^\ix{num}$ requires such a large value since the ground
state energy depends on $\lambda_\ix{i}^\ix{num}$ in leading order via
an addend proportional to $1/\sqrt{\lambda_\ix{i}^\ix{num}}$.

Obviously, the self-consistent Hartree-Fock approximation, that is
$C$-flow truncated at $m=2$, yields already quite good results with a
relative discrepancy of less than $3\%$ for $g$ up to $50$.  For the
problem at hand, this is plausible as mean-field theory can be
expected to capture reasonably well the effect of the
$x^4$-anharmonicity on low-energy properties by self-consistently
optimizing the frequency of the harmonic oscillator.  The improvement
achieved by second order self-consistent perturbation theory, that is
$C$-flow truncated at $m=4$, is rather low in comparison: the data is
typically $1$ to $2\%$ off for $e_\ix{gs}$, while $\varrho$ is
reproduced even less accurate than by the Hartree-Fock approximation
for $g\gtrsim 20$.  The results from internally closed $C$-flow are
significantly better, with relative deviations below $1\%$ for both
observables.  The idea of internally closing the set of differential
equations by expressing $\overline{\Phi}{}^{(3)}$ in terms of
$\overline{\Phi}{}^{(2)}$ is thus the most successful $C$-flow scheme
investigated here. However, the price to pay is a large computational
effort $\sim m_\ix{len}^4$ for the computation of the set of differential equations in each step of the flow. Additionally, it
is unclear whether this approach can be systematically generalized to
higher orders. Therefore, it is not possible to compare results from
truncations at different orders in order to assess the quality of a
given approximation, a procedure often used in fRG studies
\cite{Met12}. Anyhow, hypothetical higher orders of internally closed
$C$-flow seem out of reach for numerical computations as the scheme implemented here is already numerically quite demanding.

The question emerges whether one can think of an RG scheme that better
fits the 2PI nature of the general approach. One possible answer is
the $U$-flow that shall be discussed in Sect.~\ref{sec:U-flow}.
Sect.~\ref{sec:Bethe-Salpeter} provides some preliminaries for this
discussion.


\section{Pair propagator and Bethe-Salpeter equation}

\label{sec:Bethe-Salpeter}

In the discussion of the $U$-flow scheme below, some notions related
to four-point functions and the dot-product introduced above will be
useful. They are provided in this section which can be considered as a
sequel to Sect.~\ref{sec:gen-func}.

From the non-interacting functional
$W_0[J]=-\frac{\zeta}{2}\tr\ln(-C^{-1}-J)$ follows $G_0[J] = -
W_0^{(1)}[J] = (C^{-1}+J)^{-1}$. From $W^{(2)}_0[J] = -\delta
G_0[J]/\delta J$ follows $W^{(2)}_0\big[J_0[G]\big] = \Pi[G]$, with
the pair propagator
\begin{equation}
  \fl
  \Pi_{\gamma_1 \gamma_2}[G]
  = -\frac{\delta G_{\gamma_2}}{\delta G^{-1}_{\gamma_1}} 
  = G_{\alpha_2 \alpha_3} \frac{\delta G^{-1}_{\alpha_3
      \alpha'_3}}{\delta G^{-1}_{\alpha_1 \alpha'_1}} G_{\alpha'_3
    \alpha'_2} 
  = G_{\alpha'_1 \alpha_2} G_{\alpha_1 \alpha'_2} + \zeta G_{\alpha_1
    \alpha_2} G_{\alpha'_1 \alpha'_2}.
\end{equation}
In case of a particle number conserving system, $\Pi$ denotes either a
particle-hole or a particle-particle propagator, depending on the
charge indices. For example the charge indices $c_1 = -c'_1 = c_2  =
-c'_2$ render $\Pi_{\gamma_1 \gamma_2}$ a particle-hole propagator,
while the charge indices $c_1 = c'_1 = - c_2  = -c'_2$ lead to a
particle-particle propagator.

In the general interacting case, one has
\begin{equation}
  W^{(2)} \cdot \Gamma^{(2)} = \frac{\delta W^{(1)}}{\delta J} \cdot
  \frac{\delta \Gamma^{(1)}}{\delta G} = \frac{\delta G}{\delta J}
  \cdot \frac{\delta J}{\delta G} = \frac{\delta J}{\delta J} = I,
\end{equation}
where
\begin{equation}
  I_{\gamma_1 \gamma_2} = \frac{\delta J_{\gamma_2}}{\delta
    J_{\gamma_1}} = \delta_{\alpha_1 \alpha_2} \delta_{\alpha'_1
    \alpha'_2} + \zeta \delta_{\alpha_1 \alpha'_2} \delta_{\alpha'_1
    \alpha_2} 
\end{equation}
is the neutral element with respect to the dot-product, $X \cdot I = I \cdot X = X$.  We conclude
\begin{equation}
  W^{(2)} = \big(\Gamma^{(2)}\big)^\inv,
\end{equation}
and in particular $\Gamma_0^{(2)}=\Pi^\inv$.  Here, we introduced the dot-product inverse of four-point functions that satisfies
\begin{equation}
  \big(X \cdot X^\inv\big)_{\gamma_1 \gamma_2} 
  = \frac{1}{2}\sum_{\gamma_3} X_{\gamma_1 \gamma_3} X^\inv_{\gamma_3
    \gamma_2} 
  =  I_{\gamma_1 \gamma_2}.
\end{equation}
An example is
\begin{equation}
  \left(\Pi^\inv\right)_{\gamma_1 \gamma_2}[G] = G^{-1}_{\alpha'_1
    \alpha_2} G^{-1}_{\alpha_1 \alpha'_2} + \zeta G^{-1}_{\alpha_1
    \alpha_2} G^{-1}_{\alpha'_1 \alpha'_2 },
\end{equation}
where $G^{-1}$ denotes the regular inverse with respect to the
$\alpha$-indices.  Now, $\Gamma^{(2)} = \Phi^{(2)} + \Gamma_0^{(2)} =
\Phi^{(2)} + \Pi^\inv$ leads to the Bethe-Salpeter equation
\begin{equation}
\label{eq:Bethe-Salpeter}
\fl
  W^{(2)} =  \big(\Pi^\inv + \Phi^{(2)}\big)^\inv = \Pi - \Pi \cdot
  \Phi^{(2)} \cdot \Pi \pm \dots = \Pi - \Pi \cdot \Phi^{(2)} \cdot W^{(2)}.
\end{equation}


\section{$U$-flow}
\label{sec:U-flow}


\subsection{Flow parameter, flow equations, and initial conditions for
  the plain $U$-flow}
\label{sec:U-flow-flow-param}

In 1PI fRG, the flow parameter is typically introduced in the
one-particle propagator.  In $C$-flow, this procedure was carried over
to the 2PI case.  We next explore the consequences of introducing the
flow parameter into the interaction, $U \to U_\lambda$, as done in
Ref.~\cite{Dup14}. This method will be denoted as ``$U$-flow''. We set
$U_{\lambda_\ix{i}}=0$ at the beginning of the flow, such that the
flow starts at the non-interacting model. At the end of the flow,
$U_{\lambda_\ix{f}}=U$ restores the original interacting problem. We
maintain the full index permutation (anti-)symmetry of the interaction
during all of the flow, such that $\dot U_{\alpha_1 \alpha_2 \alpha_3
  \alpha_4} = \zeta^P \dot U_{\alpha_{P1} \alpha_{P2} \alpha_{P3}
  \alpha_{P4}}$.

In Ref.~\cite{Hon04}, an ``interaction flow'' method in the framework
of 1PI fRG was studied. The authors describe that a flow parameter $C
\rightarrow \lambda C$ can be substituted to a flow parameter $U
\rightarrow \lambda^2 U$ by rescaling the fields. One could jump to
the conclusion that there is in principle no difference between
$C$-flow and $U$-flow, at least for a multiplicative flow parameter.
The rescaling of the fields, however, leads to a $\lambda$-dependent
rescaling of all Green and vertex functions. Thus, the structure of the flow
equations for the non-rescaled vertex functions differs for the two
flow parameters and suggests different truncation procedures.

It is conceivable to combine the $C$-flow and $U$-flow schemes by
introducing a flow parameter into both, $C$ and $U$, compare
Ref.~\cite{Dup14} and also \cite{Kem13}. Here, we discuss only the
pure $U$-flow. In a combined scheme, the flow equations will include
additionally the terms derived for the $C$-flow in
Sect.~\ref{sec:C-flow_flow-parameter}.

Possible infrared divergencies of perturbation theory ought to be
regularized by the chosen flow parameter. This can indeed be achieved
in the framework of the $U$-flow. It has been shown e.g. in the
context of partial bosonization for one-dimensional models with
interactions that involve only small momentum transfers that a cutoff
in this transfer regularizes all infrared divergencies \cite{Sch05} (see also \cite{Dup14}).

$U_\lambda$ induces a $\lambda$-dependence of the action and of the
functionals. We now derive the flow equations for the $\overline
\Phi{}^{(n)}_\lambda$. We refer to this scheme as ``plain $U$-flow'' in
order to distinguish it from the modified scheme that is proposed in
Ref.~\cite{Dup14} and which we describe in Sect.~\ref{sec:modU}.  The
flow of the functionals is given by
\begin{eqnarray}
  \dot Z
  &= - \frac{1}{3!} \Tr \dot U \cdot \frac{\delta^2 Z}{\delta J
    \delta J},
  \\
  \dot W
  &= \frac{\dot Z}{Z} = - \frac{1}{3!} \left[\Tr \left(\dot U \cdot
      W^{(2)}\right) + W^{(1)} \cdot \dot U \cdot W^{(1)}\right], 
  \\
  \dot \Gamma
  &= - \dot W|_{J[G]} = \frac{1}{3!} \left[\Tr \left(\dot U \cdot
      W^{(2)}\right) + G \cdot \dot U \cdot G\right]. 
\end{eqnarray}
In the last expression, $W^{(2)}\big[J[G]\big]= \left(\Gamma^{(2)}[G]\right)^\inv$.  We used the ``double-index'' trace
\begin{equation}
  \Tr X = \frac{1}{2} \sum_\gamma X_{\gamma \gamma},
\end{equation}
compare Ref.~\cite{Dup14}, which is to be distinguished from the ``single-index'' trace $\tr Y = \sum_{\alpha} Y_{\alpha \alpha}$. As the non-interacting effective action $\Gamma_0$ does not depend on $U$, the flow of the Luttinger-Ward functional $\Phi=\Gamma-\Gamma_0$ is given by
\begin{equation}
  \label{eq:U-flow-Phi}
  \dot \Phi = \dot \Gamma = \frac{1}{3!} \Tr \dot U \cdot
  \left(W^{(2)} + \frac{\Pi}{2} \right),
\end{equation}
where $W^{(2)} = \left(\Pi^\inv + \Phi^{(2)}\right)^\inv$.

In order to compute the flow of the $\overline \Phi{}^{(n)}$ given by $\dot{\overline \Phi}{}^{(n)} = \overline{\dot \Phi}{}^{(n)} + \overline \Phi{}^{(n+1)} \cdot \dot{\overline G}$, we need expressions for $\dot{\overline G}$ and $\dot \Phi^{(n)}$. For $\dot{\overline G}$, we use Eq.~(\ref{eq:G-bar-dot}) for the practical application to the anharmonic oscillator. However, for gaining general analytical insights, a different but equivalent expression turns out to be useful: from $0 =\dot {\overline \Gamma}{}^{(1)} = \overline{\dot \Gamma}{}^{(1)} + \overline \Gamma{}^{(2)} \cdot \dot {\overline G}$ follows
\begin{equation}
  \label{eq:G-bar-dot-alt}
  \dot {\overline G} 
  = - \big( \overline{\Gamma}{}^{(2)}\big)^\inv \cdot \overline {\dot \Gamma}{}^{(1)}
  = - \overline W{}^{(2)} \cdot \overline {\dot \Phi}{}^{(1)}.
\end{equation}
For $\dot \Phi^{(1)} = \delta \dot \Phi / \delta G$, we find
\begin{eqnarray}
  \fl
  \dot \Phi^{(1)}
  &=
  \frac{1}{3!} \Tr \dot U \cdot
  \left(\frac{\delta W^{(2)}}{\delta G} + \frac{1}{2}\frac{\delta
      \Pi}{\delta G} \right)
  \\
  \fl
  &=
  \frac{1}{3!} \Tr \dot U \cdot \left[ W^{(2)} \cdot \left( \Pi^\inv
      \cdot \frac{\delta \Pi}{\delta G} \cdot \Pi^\inv  - \Phi^{(3)}
    \right) \cdot W^{(2)} + \frac{1}{2} \frac{\delta \Pi}{\delta G}
  \right];
\end{eqnarray}
writing indices explicitly, this means
\begin{eqnarray}
 \label{eq:MUF_Phi1_flow_w_indices}
  \dot \Phi^{(1)}_{\gamma_1}
  &
  = \frac{1}{3} \sum_{\gamma_2} \left(\Pi^\inv \cdot W^{(2)}
    \cdot \dot U  \cdot W^{(2)} \cdot \Pi^\inv +  \frac{\dot U}{2}
  \right)_{\alpha_1 \alpha_2 \alpha'_2 \alpha'_1} G_{\gamma_2}  
  \nonumber
  \\
  &\quad
  - \frac{1}{4!} \sum_{\gamma_2 \gamma_3} \Phi^{(3)}_{\gamma_1\gamma_2
    \gamma_3}  \left(W^{(2)} \cdot \dot U \cdot 
    W^{(2)} \right)_{\gamma_3 \gamma_2}. 
\end{eqnarray}
This yields  explicit expressions for $\dot{\overline G} = -
\overline W{}^{(2)} \cdot \overline {\dot \Phi}{}^{(1)}$, and
\begin{equation}
  \label{eq:U-flow-Phi-bar}
  \dot{\overline \Phi}
  =
  \overline{\dot \Phi} + \overline \Phi{}^{(1)} \cdot \dot{\overline G}
  =
  \overline{\dot \Phi} - \overline \Phi{}^{(1)} \cdot \overline W{}^{(2)}
  \cdot \overline {\dot \Phi}{}^{(1)},
\end{equation}
as well as
\begin{equation}
  \label{eq:U-flow-Phi1-bar}
  \fl
  \dot{\overline \Phi}{}^{(1)}
  =
  \overline{\dot \Phi}{}^{(1)} + \overline \Phi{}^{(2)} \cdot  \dot{\overline G}
  =
  \left( I - \overline \Phi{}^{(2)} \cdot \overline W{}^{(2)} \right) \cdot
  \overline{\dot \Phi}{}^{(1)}
  =
  \left( I + \overline \Phi{}^{(2)} \cdot \overline \Pi \right)^\inv\cdot
  \overline{\dot \Phi}{}^{(1)}.
\end{equation}
In order to evaluate the right-hand-side of
Eqs.~(\ref{eq:U-flow-Phi-bar}) and~(\ref{eq:U-flow-Phi1-bar}), we need
to know $\overline \Phi{}^{(1)}$, $\overline \Phi{}^{(2)}$ [which also
allows to compute $\overline W{}^{(2)} = \big(\overline \Pi^\inv+
\overline \Phi{}^{(2)}\big)^\inv$], and $\overline \Phi{}^{(3)}$
[cf. Eq.~(\ref{eq:MUF_Phi1_flow_w_indices})]. This is the onset of an
infinite hierarchy of coupled flow equations, where $\dot{\overline
  \Phi}$ depends on $\overline \Phi{}^{(1)},\dots,\overline \Phi{}^{(3)}$,
and where $\dot{\overline \Phi}{}^{(n)}$ depends on $\overline
\Phi{}^{(2)},\dots,\overline\Phi{}^{(n+2)}$ for $n \ge 1$.

The starting point of the flow is given by $U_{\lambda_\ix{i}}=0$ and corresponds to the non-interacting model. Consequently, $\Phi_{\lambda_\ix{i}} = 0$ and
\begin{equation}
  \overline \Phi_{\lambda_\ix{i}} = 0, \qquad \overline
  \Phi{}^{(n)}_{\lambda_\ix{i}} = 0, \; n = 1,2,3,\dots
\end{equation}

The flow of the thermodynamic potential is given by
\begin{equation}
  \dot{\overline \Omega} 
  = \frac{\dot{\overline \Gamma}}{\beta}
  = \frac{1}{\beta} \left(\overline{\dot \Gamma} + \overline
    \Gamma^{(1)} \cdot \dot{\overline G} \right)
  = \frac{\overline{\dot \Gamma}}{\beta}
  = \frac{\overline{\dot \Phi}}{\beta}.
\end{equation}
It starts at the non-interacting potential, $\overline
\Omega_{\lambda_\ix{i}} = \Omega_0[C] = - \frac{1}{\beta} \ln
\overline Z_0$.


\subsection{Truncation of the plain $U$-flow}

In the plain $U$-flow, the lowest truncation of the hierarchy of flow
equations that leads to a flowing self-energy is at level $m=3$: one
sets $\overline \Phi{}^{(n)}_\lambda = \overline
\Phi{}^{(n)}_{\lambda_\ix{i}} = 0$, that is to its initial value, for
$n\ge 3$. The resulting flow equation $\dot{\overline \Phi}{}^{(2)} =
\overline{\dot \Phi}{}^{(2)} + \overline \Phi{}^{(3)} \cdot
\dot{\overline G} = \overline{\dot \Phi}{}^{(2)}$ reads
\begin{eqnarray}
  \fl
  \dot{\overline \Phi}{}^{(2)}
  =
  \frac{1}{3!} \Tr \dot U 
  &
  \left[ 
    2 \overline W{}^{(2)} \cdot \overline \Pi^\inv \cdot \frac{\delta
      \overline \Pi}{\delta \overline G} \cdot \left( \overline
      \Pi^\inv  \cdot \overline W{}^{(2)} \cdot \overline \Pi^\inv -
      \overline \Pi^\inv \right) \cdot \frac{\delta \overline
      \Pi}{\delta \overline G} \cdot \overline \Pi^\inv 
    \cdot \overline W{}^{(2)} \right.
  \nonumber \\
  \fl
  & \left. \phantom{[}
    + \overline W{}^{(2)} \cdot \overline \Pi^\inv \cdot \frac{\delta^2
      \overline \Pi}{\delta \overline G^2} \cdot \overline \Pi^\inv
    \cdot \overline W{}^{(2)}
    +\frac{1}{2} \frac{\delta^2 \overline \Pi}{\delta \overline G^2}
  \right]
\end{eqnarray}
and produces a $\overline \Phi{}^{(2)}$ with explicit time or frequency dependence that enters the flow of the self-energy. Solving the flow equations is then already quite involved. In order to find a simpler approximation, we use the expansion $\overline W{}^{(2)} = \overline \Pi + \mathcal{O}\left(\overline \Phi{}^{(2)}\right)$ and get
\begin{eqnarray}
  \dot{\overline \Phi}{}^{(2)}
  =
  \frac{1}{4} \Tr \dot U \cdot \frac{\delta^2 \overline \Pi}{\delta
    \overline  G^2}
  + \mathcal{O}\left(\dot U \overline \Phi{}^{(2)}\right)
  = 
  \dot U \left[1 + \mathcal{O}\left(\overline \Phi{}^{(2)}\right)
  \right].
\end{eqnarray}
This can be integrated to $\overline \Phi{}^{(2)}_\lambda = U_\lambda + \mathcal{O}(U_\lambda^2)$ which we approximate by $\overline \Phi{}^{(2)}_\lambda = U_\lambda$.  This result can be understood also as the leading perturbative approximation to $\Phi^{(2)}$, compare Eq.~(\ref{eq:Lutt-Ward-expand}). The flow equations for the self-energy and the thermodynamic potential are then
\begin{eqnarray}
  \fl
  \dot{\overline \Sigma}{}^\lambda_{\gamma_1}
  =
  -\frac{1}{6} \sum_{\gamma_2, \gamma_3} (I + U_\lambda \cdot \overline \Pi_\lambda)^\inv_{\gamma_1 \gamma_2} 
  \nonumber \\
  \times \left[(I+ U_\lambda \cdot \overline \Pi_\lambda)^\inv \cdot \dot U_\lambda \cdot (I+\overline \Pi_\lambda \cdot U_\lambda)^\inv 
    + \frac{\dot{U}_\lambda}{2} \right]_{\alpha_2 \alpha_3 \alpha'_3 \alpha'_2} \overline G^\lambda_{\gamma_3},
\end{eqnarray}
\begin{equation}
  \dot{\overline \Omega}_\lambda = \frac{1}{3! \beta} \Tr \dot U_\lambda \cdot \left[\left(\overline \Pi_\lambda^\inv + U_\lambda \right)^\inv + \frac{\overline \Pi_\lambda}{2} \right]. 
\end{equation}

Let us address the question how the solutions of these approximate flow equations compare to perturbation theory. It turns out that they comprise first and second order perturbation theory (with bare propagators),
\begin{equation}
  \overline \Sigma_{\lambda_\ix{f}} = \overline
  \Sigma_\ix{exact} + \mathcal{O}\left(U^3\right), 
  \qquad
  \overline \Omega_{\lambda_\ix{f}} = \overline
  \Omega_\ix{exact} + \mathcal{O}\left(U^3\right).
\end{equation}
For the proof, we start from the initial condition $\overline
\Sigma_{\lambda_\ix{i}} = 0$, $\overline G_{\lambda_\ix{i}} = C$, $\overline \Pi_{\lambda_\ix{i}} = \overline \Pi_0$. From $\dot {\overline \Sigma} = \mathcal{O}(\dot U)[1+\mathcal{O}(U)]$ follows $\overline \Sigma = \mathcal{O}(U)$, hence $\overline G = C + \mathcal{O}(U)$, $\overline \Pi = \overline \Pi_0 + \mathcal{O}(U)$. This leads to
\begin{eqnarray}
  \fl
  \dot{\overline \Sigma}_{\gamma_1}
  = &
  \big[-\dot U \cdot C - \dot U \cdot (C \overline \Sigma C) + U \cdot
  \overline \Pi_0 \cdot \dot U \cdot C \big]_{\gamma_1}
  \nonumber \\ 
  \fl &
  + \frac{1}{3} \sum_{\gamma_3} (U \cdot \overline \Pi_0 \cdot \dot U
  + \dot U \cdot \overline \Pi_0 \cdot U)_{\alpha_1 \alpha_3 \alpha'_3
    \alpha'_1} C_{\gamma_3}  
  + \mathcal{O}(U^2\dot U).
\end{eqnarray}
The first addend can be integrated to $(-U \cdot C)$ which is first 
order perturbation theory for $\overline \Sigma$. We insert $\overline \Sigma = -U \cdot C + \mathcal{O}(U^2)$ into the second addend, 
\begin{equation}
  - \dot U \cdot (C \overline \Sigma C)
  = \dot U \cdot \overline \Pi_0 \cdot U \cdot C + \mathcal{O}(U^2 \dot U).
\end{equation}
Together with the third addend, this can be integrated to the value of
the mean-field-like (non-2PI) second order self-energy diagram $U \cdot
\overline \Pi_0 \cdot U \cdot C$, plus third order corrections.  The
fourth addend can be integrated to
\begin{equation}
  \frac{1}{3} \sum_{\gamma_3} (U \cdot \overline \Pi_0 \cdot
  U)_{\alpha_1 \alpha_3 \alpha'_3 \alpha'_1} C_{\gamma_3}, 
\end{equation}
which is the value of the 2PI second order self-energy diagram. Thus, we recovered all diagrams up to second order, and the remaining $\mathcal{O}(U^2\dot U)$ leads to some correction of $\mathcal{O}(U^3)$.  The proof for the thermodynamic potential is similar.


\subsection{Modified $U$-flow starting at mean-field theory}
\label{sec:modU}

In Ref.~\cite{Dup14}, a slight modification of the $U$-flow is
proposed which results in a different starting point of the flow,
namely mean-field theory instead of the non-interacting problem.  The
modified $U$-flow is defined by
\begin{equation}
  \label{eq:modified-U-flow-def}
  \Phi_\lambda(\textrm{modified $U$-flow}) =
  \Phi_\lambda(\textrm{plain $U$-flow}) + \frac{1}{2} G \cdot
  (U-U_\lambda) \cdot G,
\end{equation}
where $\Phi_\lambda(\textrm{plain $U$-flow})$ refers to the scheme described in Sect.~\ref{sec:U-flow-flow-param}. For an interpretation of the modified $U$-flow, consider the expansion of the Luttinger-Ward functional given in Eq.~(\ref{eq:Lutt-Ward-expand}). In the plain $U$-flow, each $U$ in this expansion is replaced by $U_\lambda$. In the modified $U$-flow, the first order contribution $\frac{1}{2} G \cdot U \cdot G$ is excluded from this replacement. At the starting point $U_{\lambda_\ix{i}}=0$ of the flow, $\Phi_{\lambda_\ix{i}}$ in the modified scheme is given by precisely that first order contribution,
\begin{equation}
  \fl
  \Phi_{\lambda_\ix{i}} = \frac{1}{2} G \cdot U \cdot G,
  \qquad
  \Phi^{(1)}_{\lambda_\ix{i}} = -\Sigma_{\lambda_\ix{i}} = U \cdot G,
  \qquad
  \Phi^{(2)}_{\lambda_\ix{i}} = U,
  \qquad
  \Phi^{(n)}_{\lambda_\ix{i}} = 0, \; n\ge 3.
\end{equation}
As $\Sigma = -U \cdot G$ is the Hartree-Fock self-consistency equation, we see that the flow starts at mean-field theory (cf. \cite{sue06}). At the end point $U_{\lambda_\ix{f}} = U$ of the flow, the full genuine Luttinger-Ward functional is recovered. If required, the flow of the other functionals is derived from the flow of $\Phi_\lambda$. The effective action is given by $\Gamma_\lambda = \Phi_\lambda + \Gamma_0$, and $W_\lambda$ is obtained from $\Gamma_\lambda$ by a Legendre transformation.

In the same spirit, one could define more general flow schemes that
start at $n$th order self-consistent perturbation theory. In the
definition of $\Phi_\lambda$, the replacement $U\rightarrow U_\lambda$
would be applied only to contributions of order $U^{n+1}$ and higher.
Depending on the problem of interest, starting the flow with $n$th
order self-consistent perturbation theory might or might not be
advantageous; see the discussion of truncated $C$-flow for the
anharmonic oscillator above and Ref.~\cite{Whi92} for the
single-impurity Anderson model.

In the framework of the modified scheme, the functional
$\Phi(\textrm{plain $U$-flow})$ does not play the role of the
Luttinger-Ward functional. Therefore, we label it $\Psi$ from now on
such that Eq.~(\ref{eq:modified-U-flow-def}) takes the form
$\Phi_\lambda = \Psi_\lambda + \frac{1}{2} G \cdot (U-U_\lambda) \cdot
G$. For the flow equation follows $\dot \Phi = \dot \Psi -\frac{1}{2}
G \cdot \dot U \cdot G = \dot \Psi -\frac{1}{4} \Tr \dot U \cdot \Pi$. We insert $\dot \Psi$ from Eq.~(\ref{eq:U-flow-Phi}) and obtain
\begin{equation}
  \dot \Phi
  =
  \frac{1}{3!} \Tr \dot U \cdot \left[\left(\Pi^\inv +
      \Psi^{(2)} \right)^\inv -\Pi \right],
\end{equation}
\begin{eqnarray}
  \fl
  \dot \Phi^{(1)}_{\gamma_1}
  &= \!\frac{1}{3} \sum_{\gamma_2} \!\left[\!\left(I + \Psi^{(2)} \cdot \Pi \right)^{\!\inv} \!\! \cdot \dot U  
    \cdot \left(I + \Pi \cdot \Psi^{(2)} \right)^{\!\inv} \!\!
    - \dot U \right]_{\alpha_1 \alpha_2 \alpha'_2 \alpha'_1} \!\!
  G_{\gamma_2}  
  + \mathcal{O}\!\left(\Psi^{(3)} \right),
\end{eqnarray}
where $\Psi^{(2)}_\lambda = \Phi^{(2)}_\lambda + U_\lambda - U$ and $\Psi^{(3)} = \Phi^{(3)}$. On the analogy of Eqs.~(\ref{eq:U-flow-Phi-bar}) and~(\ref{eq:U-flow-Phi1-bar}) follows
\begin{equation}
  \fl
  \dot{\overline \Phi}
  =
  \overline{\dot \Phi} - \overline \Phi{}^{(1)} \cdot \left(\overline \Pi^\inv +
    \overline \Phi{}^{(2)} \right)^\inv 
  \cdot \overline {\dot \Phi}{}^{(1)},
  \qquad
  \dot{\overline \Phi}{}^{(1)}
  =
  \left( I + \overline \Phi{}^{(2)} \cdot \overline \Pi \right)^\inv\cdot
  \overline{\dot \Phi}{}^{(1)}.
\end{equation}

As for the plain $U$-flow, the flow equation for the thermodynamic potential has the form $\dot{\overline \Omega} = \overline{\dot \Phi}/\beta$. The initial value is now the mean-field potential $\overline \Omega_\ix{MF}$ indicated in Eq.~(\ref{eq:Omega-MF}).


\subsection{Truncation of the modified $U$-flow}

In the modified $U$-flow, the lowest truncation of the hierarchy of flow equations that leads to a flowing self-energy $\overline \Sigma_\lambda = - \overline \Phi{}^{(1)}_\lambda$ is at level $m=2$: set $\overline \Phi{}^{(2)}_\lambda = \overline \Phi{}^{(2)}_{\lambda_\ix{i}} = U$ and $\overline \Phi{}^{(n)}_\lambda = \overline \Phi{}^{(n)}_{\lambda_\ix{i}} = 0, n\ge 3$. The flow of the self-energy in this approximation is
\begin{eqnarray}
  \fl
  \dot{\overline \Sigma}{}^\lambda_{\gamma_1}
  =
  -\frac{1}{6} \sum_{\gamma_2, \gamma_3}
  (I + U\cdot \overline \Pi_\lambda)^\inv_{\gamma_1 \gamma_2} 
  \nonumber \\
  \times \left[(I+ U_\lambda \cdot
    \overline \Pi_\lambda)^\inv \cdot \dot U_\lambda \cdot (I+\overline
    \Pi_\lambda \cdot U_\lambda)^\inv - \dot U_\lambda \right]_{\alpha_2
    \alpha_3 \alpha'_3 \alpha'_2} \overline G{}^\lambda_{\gamma_3},
\end{eqnarray}
where $U_\lambda$ refers to the flow parameter dressed value of the interaction, while $U$ denotes the bare one. The flow equation for the thermodynamic potential $\dot{\overline \Omega} = \overline{\dot \Phi}/\beta$ reads
\begin{equation}
  \dot{\overline \Omega}_\lambda = \frac{1}{3! \beta} \Tr 
  \dot U_\lambda \cdot \left[\left(\overline \Pi_\lambda^\inv +
      U_\lambda \right)^\inv - \overline \Pi_\lambda  \right].
\end{equation}

It can be shown that the solutions of these approximate flow equations comprise all diagrams contributing to self-consistent Hartree-Fock (mean-field) and to second order perturbation theory with mean-field propagators,
\begin{equation}
  \overline \Sigma_{\lambda_\ix{f}} = \overline
  \Sigma_\ix{exact} + \mathcal{O}\left(U^3 \overline G_\ix{MF}^5 \right), 
  \qquad
  \overline \Omega_{\lambda_\ix{f}} = \overline
  \Omega_\ix{exact} + \mathcal{O}\left(U^3 \overline G_\ix{MF}^6 \right).
\end{equation}
For the proof, we start from the initial condition $\overline
\Sigma_{\lambda_\ix{i}} = \overline \Sigma_\ix{MF}$, $\overline
G_{\lambda_\ix{i}} = \overline G_\ix{MF}$, $\overline
\Pi_{\lambda_\ix{i}} = \overline \Pi_\ix{MF}$. From $\dot {\overline
  \Sigma} = \mathcal{O}\big(U \dot U \overline G^3\big)$ follows $\overline \Sigma_\lambda =
\overline \Sigma_\ix{MF}+ \mathcal{O}\big(U^2 \overline G_\ix{MF}^3\big)$, hence $\overline
G_\lambda = \overline G_\ix{MF} + \mathcal{O}\big(U^2 \overline
G_\ix{MF}^5\big)$, $\overline \Pi_\lambda = \overline \Pi_\ix{MF} +
\mathcal{O}\big(U^2 \overline G_\ix{MF}^6\big)$. Therefore,
\begin{eqnarray}
  \fl
  \dot{\overline \Sigma}{}^\lambda_{\gamma_1}
  &=
  \frac{1}{3} \sum_{\gamma_3}
  \left(U_\lambda \cdot \overline \Pi_\ix{MF} \cdot \dot U_\lambda 
    + \dot U_\lambda \cdot \overline \Pi_\ix{MF} \cdot U_\lambda
  \right)_{\alpha_1 \alpha_3 \alpha'_3 \alpha'_1} \overline
  G_{\ix{MF},\gamma_3} 
  + \mathcal{O}\left(U^2 \dot U \overline G_\ix{MF}^5 \right).
\end{eqnarray}
Integration from $\lambda_\ix{i}$ to $\lambda_\ix{f}$ yields
\begin{equation}
  \overline \Sigma_{\lambda_\ix{f}} 
  = \overline \Sigma_\ix{MF} 
  +  \frac{1}{3} \sum_{\gamma_3}
  \left(U \cdot \overline \Pi_\ix{MF} \cdot U \right)_{\alpha_1
    \alpha_3 \alpha'_3 \alpha'_1} \overline G_{\ix{MF},\gamma_3} 
  + \mathcal{O}\left(U^3 \overline G_\ix{MF}^5 \right).
\end{equation}
The second addend on the right-hand-side is indeed the 2PI second
order self-energy diagram with mean-field propagators.  The proof for
$\overline \Omega$ can be performed
analogously.


\subsection{Application to the anharmonic oscillator}

For a practical application of plain and modified $U$-flow, we use Eq.~(\ref{eq:G-bar-dot}) for the calculation of $\dot{\overline{G}}$. In plain $U$-flow, one obtains for the self-energy (still for general $\gamma$-indices)
\begin{eqnarray}
  \fl
  \dot{\overline \Sigma}{}^\lambda_{\gamma_1}
  = &
  -\frac{1}{3} \sum_{\gamma_2} \left[\left(I+ U_\lambda \cdot \overline \Pi_\lambda\right)^\inv \cdot \dot U_\lambda 
  \cdot \left(I+\overline \Pi_\lambda \cdot U_\lambda\right)^\inv + \frac{\dot{U}_\lambda}{2} \right]_{\alpha_1 \alpha_2 \alpha'_2 \alpha'_1} \overline G{}^\lambda_{\gamma_2}
  \nonumber \\
  \fl & -\frac{1}{2} \sum_{\gamma_2} U_{\lambda, \gamma_1 \gamma_2} \dot{\overline{G}}{}^\lambda_{\gamma_2}.
\end{eqnarray}
Introducing the quantity $\overline{\Upsilon}_\lambda=I-\left(I+ U_\lambda \cdot \overline{\Pi}_\lambda\right)^\inv$, one finds
\begin{eqnarray}
  \fl
  \dot{\overline \Sigma}{}^\lambda_{\gamma_1}
  =
  \!-\frac{1}{3} \sum_{\gamma_2}\!\left[ \!\!\left(\!\overline{\Upsilon}_\lambda \!\cdot\! \dot U_\lambda \!\cdot\! \overline{\Upsilon}_\lambda^\intercal
  \!-\! \overline{\Upsilon}_\lambda \!\cdot\! \dot{U}_\lambda \!-\! \dot{U}_\lambda \!\cdot\! \overline{\Upsilon}_\lambda^\intercal 
  \!+\! \frac{3}{2} \dot{U}_\lambda \!\right)_{\!\!\alpha_1 \alpha_2 \alpha'_2 \alpha'_1} \!\!\overline G^\lambda_{\gamma_2}
  +\frac{3}{2} U_{\lambda, \gamma_1 \gamma_2} \dot{\overline{G}}{}^\lambda_{\gamma_2} \right]\!\!,
\end{eqnarray}
\begin{equation}
  \dot{\overline \Omega} = -\frac{1}{3! \beta} \Tr \dot U_\lambda \cdot \overline{\Pi}_\lambda \cdot \left[\overline{\Upsilon}_\lambda-\frac{3}{2} I \right]. 
\end{equation}
On the other hand, one finds in the modified $U$-flow that
\begin{eqnarray}
  \fl
  \dot{\overline \Sigma}{}^\lambda_{\gamma_1}
  =
  \!-\frac{1}{3} \sum_{\gamma_2} \left[\overline{\Upsilon}_\lambda \!\cdot\! \dot U_\lambda \!\cdot\! \overline{\Upsilon}_\lambda^\intercal - \overline{\Upsilon}_\lambda \!\cdot\! \dot{U}_\lambda 
  - \dot{U}_\lambda \!\cdot\! \overline{\Upsilon}_\lambda^\intercal \right]_{\alpha_1 \alpha_2 \alpha'_2 \alpha'_1} \overline G^\lambda_{\gamma_2}
  -\frac{1}{2} \sum_{\gamma_2} U_{\gamma_1 \gamma_2} \dot{\overline{G}}{}^\lambda_{\gamma_2},
\end{eqnarray}
\begin{equation}
  \dot{\overline \Omega} = -\frac{1}{3! \beta} \Tr \dot U_\lambda \cdot \overline{\Pi}_\lambda \cdot \overline{\Upsilon}_\lambda. 
\end{equation}
Note that due to $(\overline{\Upsilon}^\intercal_\lambda)_{\gamma_1,\gamma_2}=\overline{\Upsilon}_{\lambda;\gamma_2,\gamma_1}$, all occurrences of $\overline{\Upsilon}^\intercal$ can generally be replaced by $\overline{\Upsilon}$. This shall be done below.

Now, we again consider the specific problem of the anharmonic
oscillator [i.e. $\zeta=+1$, $\gamma=(\tau,\tau^\prime)$]. The above equations shall be Fourier transformed according to Eqs.~(\ref{eq:FT_1}) and~(\ref{eq:FT_2}). Note that $\overline{\Upsilon}_\lambda^{(\intercal)}$ is neither propagator nor vertex but rather $\sim I$. Suitable Fourier transformation definitions are
\begin{eqnarray}
\fl \overline{\Upsilon}_{n_1 n_1^\prime n_2 n_2^\prime} &=\!\int_{\tau_1}\! \dots \int_{\tau_2^\prime} e^{+i \omega_{n_1} \tau_1 + i \omega_{n_1^\prime} \tau_1^\prime-i \omega_{n_2} \tau_2 - i \omega_{n_2^\prime} \tau_2^\prime} \overline{\Upsilon}\!\left(\tau_1,\tau_1^\prime,\tau_2,\tau_2^\prime\right) ,
\\ \fl \overline{\Upsilon}_{n_1 n_1^\prime n_2 n_2^\prime}^\intercal &=\!\int_{\tau_1}\! \dots \int_{\tau_2^\prime} e^{-i \omega_{n_1} \tau_1 - i \omega_{n_1^\prime} \tau_1^\prime+i \omega_{n_2} \tau_2 + i \omega_{n_2^\prime} \tau_2^\prime} \overline{\Upsilon}^\intercal\left(\tau_1,\tau_1^\prime,\tau_2,\tau_2^\prime\right) .
\end{eqnarray}

Let us specify the flow parameter dependence of $U_\lambda$ via a function $f_\lambda \in \mathbb{R}$ with $f_{\lambda_\textrm{\tiny i}}=0$ and $f_{\lambda_\textrm{\tiny f}}=1$ such that 
\begin{equation}
 U_\lambda(\tau_1,\tau_1^\prime,\tau_2,\tau_2^\prime)=f_\lambda g \delta(\tau_1=\tau_1^\prime=\tau_2=\tau_2^\prime).
\end{equation}
In practical calculations, $f_\lambda=\lambda$ with $\lambda_\textrm{\tiny i}=0$ and $\lambda_\textrm{\tiny f}=1$ is used. The choice of $U_\lambda$ makes it possible to show that 
\begin{eqnarray}
\fl \overline{\Upsilon}_{\lambda;n_1 n_1^\prime n_2} &=\left[1+\frac{f_\lambda g}{2 \beta} \sum_n \overline{G}{}^\lambda_{-n} \overline{G}{}^\lambda_{n-n_1-n_1^\prime} \right]^{-1} f_\lambda g \overline{G}{}^\lambda_{-n_2} \overline{G}{}^\lambda_{n_2-n_1-n_1^\prime} ,
\\ \fl \overline{\Upsilon}^\intercal_{\lambda;n_1 n_1^\prime n_2} &=f_\lambda g \overline{G}{}^\lambda_{n_1} \overline{G}{}^\lambda_{n_1^\prime} \left[1+\frac{f_\lambda g}{2 \beta} \sum_n \overline{G}{}^\lambda_n \overline{G}{}^\lambda_{n_1+n_1^\prime-n} \right]^{-1} .
\end{eqnarray}
As expected from $(\overline{\Upsilon}^\intercal_\lambda)_{\gamma_1,\gamma_2}=\overline{\Upsilon}_{\lambda;\gamma_2,\gamma_1}$, one finds $\overline{\Upsilon}^\intercal_{\lambda;n_2,n_1+n_1^\prime-n_2,n_1}=\overline{\Upsilon}_{\lambda;n_1,n_1^\prime,n_2}$. From now on, we will only use $\overline{\Upsilon}_\lambda$. It turns out that the third index of $\overline{\Upsilon}_\lambda$ is always summed over independently. This sum depends only on the sum of the first to indices. Thus, we introduce the definitions 
\begin{eqnarray}
\fl \widetilde{\Upsilon}_{\lambda;n_1+n_1^\prime}&=\frac{1}{\beta} \sum_{n_2} \overline{\Upsilon}_{\lambda;n_1 n_1^\prime n_2} =\left[1+\frac{f_\lambda g}{2} \left( \overline{G}\overline{G} \right)_{n_1+n_1^\prime} \right]^{-1} f_\lambda g \left( \overline{G}\overline{G} \right)_{n_1+n_1^\prime} ,
\\ \fl \left( \overline{G}\overline{G} \right)_N&=\frac{1}{\beta} \sum_n \overline{G}_{-n} \overline{G}_{n-N} .
\end{eqnarray}

With these findings, the flow equations in plain $U$-flow become
\begin{equation}
\dot{\overline{\Omega}}_\lambda=-\frac{\dot{f}_\lambda g}{4! \beta^2} \sum_{n_1 n_2} \overline{G}{}^\lambda_{n_1-n_2} \overline{G}^\lambda_{n_2} \widetilde{\Upsilon}_{\lambda;n_1}+\frac{\dot{f}_\lambda g}{8} \left( \frac{1}{\beta} \sum_n \overline{G}^\lambda_n \right)^2 ,
\end{equation}
\begin{equation}
\fl \dot{\overline{\Sigma}}{}^\lambda_{n_1}=- \frac{\dot{f}_\lambda g}{12 \beta} \sum_{n} \!\widetilde{\Upsilon}_{\lambda;n}^2 \overline{G}^\lambda_{n-n_1}\!+\!\frac{\dot{f}_\lambda g}{3 \beta} \sum_{n} \!\widetilde{\Upsilon}_{\lambda;n} \overline{G}^\lambda_{n-n_1} \!-\!\frac{\dot{f}_\lambda g}{2\beta}  \sum_n \!\overline{G}^\lambda_n -\frac{f_\lambda g}{2\beta} \sum_n \dot{\overline{G}}{}^\lambda_n.
\end{equation}
The last addend in the second equation depends on $\dot{\overline{\Sigma}}{}^\lambda_{n}$ and makes the equation a self-consistent one. However, this self-consistency problem can be solved explicitly as the addend does not depend on $n_1$ (i.e. it is constant with respect to the external frequency). 
With the ansatz
$\dot{\overline{\Sigma}}{}^\lambda_{n_1}=\overline{D}{}^\lambda_{n_1}+\overline{C}{}^\lambda$ where $\overline{D}{}^\lambda_{n_1}$ denotes the first three addends and $\overline{C}{}^\lambda$ the last addend, one finds:
\begin{eqnarray}
\overline{C}{}^\lambda=- \left[1+\frac{f_\lambda g}{2\beta} \sum_{n_2} \left(\overline{G}{}^\lambda_{n_2}\right)^2\right]^{-1} \frac{f_\lambda g}{2\beta} \sum_{n_2} \left(\overline{G}{}^\lambda_{n_2}\right)^2 \overline{D}{}^\lambda_{-n_2} ,
\end{eqnarray}
With the same ansatz
$\dot{\overline{\Sigma}}{}^\lambda_{n_1}=\overline{D}{}^\lambda_{n_1}+\overline{C}{}^\lambda$, one finds for the modified $U$-flow:
\begin{eqnarray}
\dot{\overline{\Omega}}_\lambda=-\frac{\dot{f}_\lambda g}{4! \beta^2} \sum_{n_1 n_2} \widetilde{\Upsilon}_{\lambda;n_1} \overline{G}{}^\lambda_{n_1-n_2} \overline{G}{}^\lambda_{n_2}  ,
\\ \overline{D}{}^\lambda_{n_1}=-\frac{\dot{f}_\lambda g}{12 \beta} \sum_{n} \widetilde{\Upsilon}_{\lambda;n}^2  \overline{G}{}^\lambda_{n-n_1} + \frac{\dot{f}_\lambda g}{3 \beta} \sum_{n} \widetilde{\Upsilon}_{\lambda;n} \overline{G}{}^\lambda_{n-n_1},
\\ \overline{C}{}^\lambda=- \left[1+\frac{g}{2\beta} \sum_{n_2} \left(\overline{G}{}^\lambda_{n_2}\right)^2\right]^{-1} \frac{g}{2\beta} \sum_{n_2} \left(\overline{G}{}^\lambda_{n_2}\right)^2 \overline{D}{}^\lambda_{-n_2} .
\end{eqnarray}
In both schemes, a non-trivial frequency dependence is generated in the self-energy in these lowest-order truncations (typically, this is different in 1PI fRG schemes). Note that the following symmetries hold: $\overline{G}_{n}=\overline{G}_{-n}$, $\left( \overline{G}\overline{G} \right)_N=\left( \overline{G}\overline{G} \right)_{-N}$, $\widetilde{\Upsilon}_{N}=\widetilde{\Upsilon}_{-N}$ and $\overline{D}_{n}=\overline{D}_{-n}$.

Now, the $\beta \to \infty$ (i.e. $T \to 0$) limit can be performed. Then, the initial conditions in plain $U$-flow are
\begin{equation}
\overline{\Omega}_{\lambda_\textrm{\tiny i}}=\frac{\omega_\textrm{\scriptsize G}}{2} , \qquad \overline{\Sigma}_{\lambda_\textrm{\tiny i}}(\nu)=0 .
\end{equation}
The initial conditions in modified $U$-flow are
\begin{equation}
\fl \overline{\Omega}_{\lambda_\textrm{\tiny i}}= \frac{1}{2} \sqrt{\omega_\textrm{\scriptsize G}^2+\overline{\Sigma}_\textrm{\scriptsize MF}} -\frac{1}{2 g} \overline{\Sigma}_\textrm{\scriptsize MF}^2 , \qquad \overline{\Sigma}_{\lambda_\textrm{\tiny i}}(\nu)=\overline{\Sigma}_\textrm{\scriptsize MF}=\frac{g}{4} \frac{1}{\sqrt{ \omega_\textrm{\scriptsize G}^2+\overline{\Sigma}_\textrm{\scriptsize MF}}} .
\end{equation}


\subsection{Numerical results for the anharmonic oscillator}

\begin{figure}
 \includegraphics[width=0.5\textwidth]{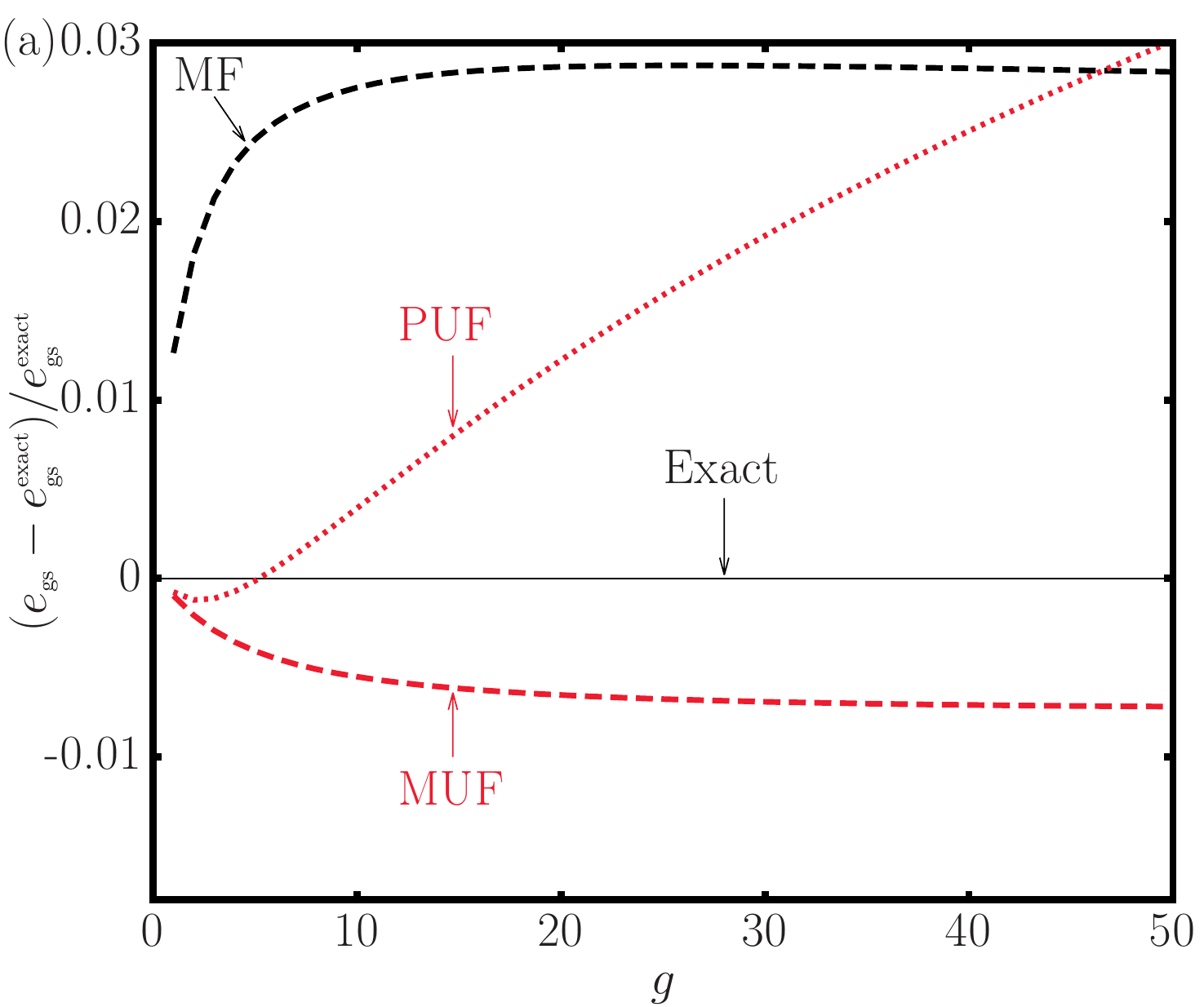}
 \includegraphics[width=0.5\textwidth]{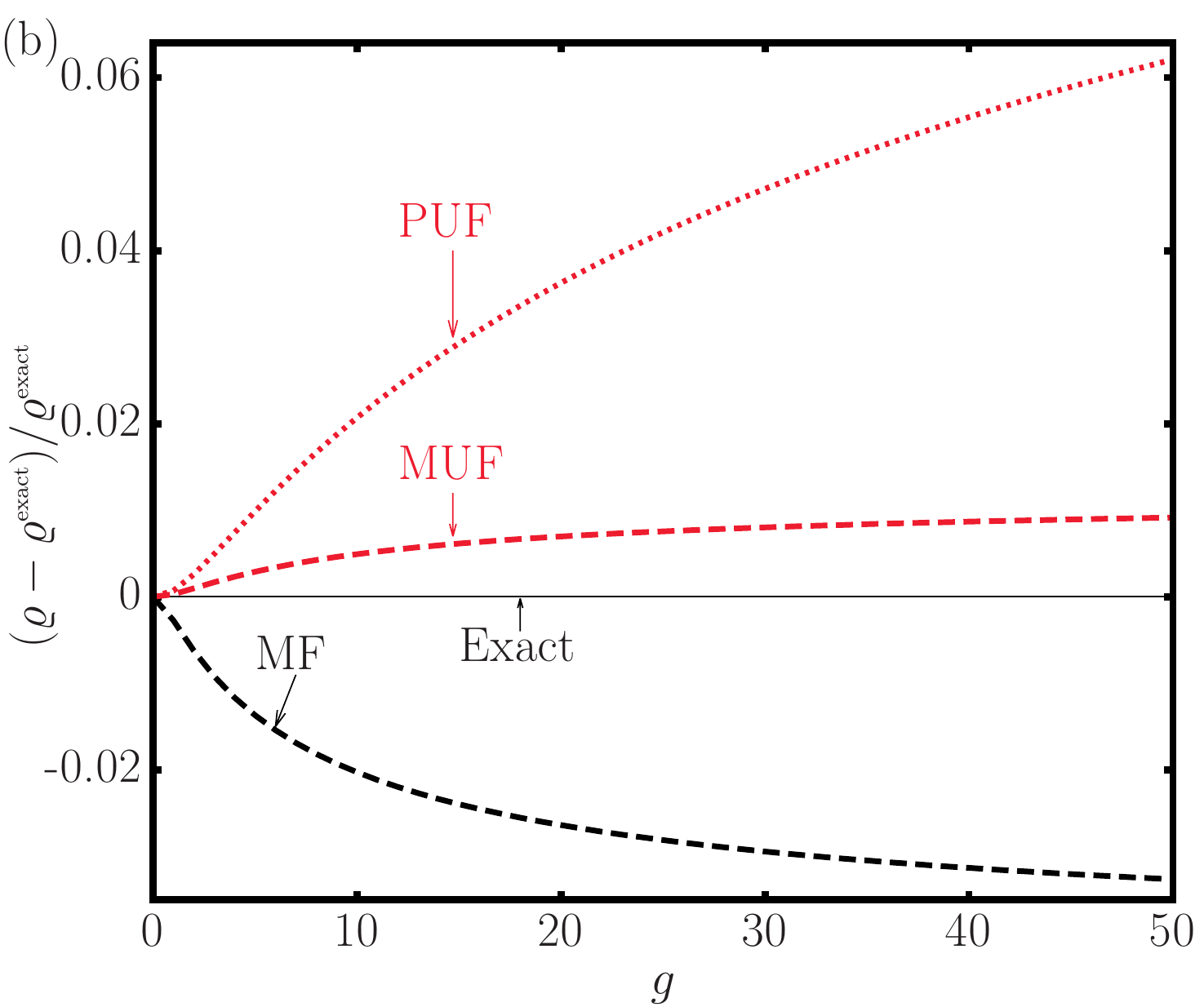}
 \caption{\label{fig:Paper_QAOsc_U-flow} 
 Numerical results for (a) the ground state energy and (b) the density as a function of the anharmonicity $g$.
 The relative difference to the exact results is shown. 
 The lines present results obtained with 
 the plain $U$-flow (dotted; PUF)  and the modified $U$-flow (light [red] dashed; MUF).
 The $U$-flow data is calculated 
 with $m_\textrm{\scriptsize len}=180$. 
The mean-field (MF) data are given for comparison.}
 \end{figure}

We implemented the zero temperature $U$-flow equations making use of the non-uniform frequency grid introduced
in \ref{App:Num_implementation}. In both schemes, the number of 
equations scales $\sim m_\textrm{\scriptsize len}$ and the effort for 
computing the right-hand-side is also $\sim m_\ix{len}$; the effort for the calculation 
of the set of differential equations in each step of the flow is thus $\sim m_\ix{len}^2$ (like 
in regular $C$-flow at truncation order $m=4$). Reaching numerical convergence with respect 
to $m_\textrm{\scriptsize len}$ is thus 
not an issue 
as it can be tuned to rather large values. Fig.~\ref{fig:Paper_QAOsc_U-flow} shows 
converged numerical data for $m_\textrm{\scriptsize len}=180$.

The modified $U$-flow curve is a lot closer to the exact result than the plain $U$-flow curve. In fact, it reproduces the exact results within $1\%$ and can thus compete with the internally closed $C$-flow method. Recall, however, that the effort to calculate the set of differential equations in each step of the flow scales $\sim m_\textrm{\scriptsize len}^2$ in the modified $U$-flow while it scales $\sim m_\textrm{\scriptsize len}^4$ in the internally closed $C$-flow. There is an intuitive reason for the success of the modified $U$-flow: 
the initial condition corresponds to the Hartree-Fock approximation which reproduces the exact results quite 
well already ($<3\%$). Thus, much of the interacting physics is captured 
initially and the flow only has to take care of the missing part. 
Apparently, starting the flow from the non-interacting system is a significantly more 
unfavourable initial condition which explains the less successful results of plain $U$-flow. 
This difference between the $U$-flow schemes can also be understood from the respective discussions 
of the degree of perturbation theory contained: 
Both schemes comprise second order perturbation 
theory but plain $U$-flow does so with the bare propagator while 
modified $U$-flow does so with the mean-field propagator.


\section{DFT-fRG}
\label{sec:DFT-fRG}

A method similar to $U$-flow 
was proposed in Refs.~\cite{Pol02,swe04} and applied to the quantum anharmonic oscillator 
in Ref.~\cite{Kem13}. 
Instead of working with an external source depending 
on two indices [like in Eq.~(\ref{eq:Z_of_J})], a purely diagonal 
one 
$J_{\alpha_1 \alpha_2}=\delta_{\alpha_1 \alpha_2} J_{\alpha_1}$ is introduced. 
Thus, the formal starting point of the considerations is not a 2PI functional but rather a two-particle 
point-irreducible (``2PPI'') one. This term is introduced in Ref.~\cite{Bra12} with the following descriptive explanation: 
a 2PPI diagram is a 1PI diagram that cannot be split into two 
by cutting two internal lines attached to the same vertex. 
This implies differences in the derivation of the flow equations and in their structure. 
In particular, it turns out that the Luttinger-Ward functional cannot be computed reasonably because the 
relation $G=-\frac{\delta W}{\delta J}$ cannot be solved explicitly for $J$ any more. The manifest remaining option 
is to work with the functional $\Gamma$ and its moments. Furthermore, the diagonality of $J$ causes $G$ to 
be the density of the physical system.
According to the discussion of Sect.~\ref{sec:compOm}
the ground state energy can be expressed through the functional $\Gamma[G]$. 
Therefore, one works with an energy functional that depends on the density. 
This brings about a close relation to density functional theory \cite{Grossbook} (in the plain Hohenberg-Kohn sense) 
and motivates us to use the term DFT-fRG. The physical state indicated by a bar $\overline{\phantom{G}}$ is 
defined as above, namely by the minimum of the energy functional. Thus, this method yields the minimum 
of the energy and the corresponding density of the interacting system at the end of the flow. In this sense, 
DFT-fRG is a way of conducting density functional theory calculations alternative to the Kohn-Sham 
idea \cite{Grossbook}. 
Truncated DFT-fRG can even be used to derive an approximation to the energy functional $\Gamma[G]$ of the 
interacting system. It remains to be seen if this approximation---which has the structure 
of a Taylor expansion around the ground state density, see below---leads to useful results for fermionic 
many-body problems.

Introducing the flow parameter simply via $U \to \lambda U$, $\lambda_\textrm{\scriptsize i}=0$ 
and $\lambda_\textrm{\scriptsize f}=1$, flow equations can be derived for $\overline{\Gamma}{}^{(n)}$ 
without further conceptual complications. However, the initial conditions $\overline{\Gamma}{}^{(n)}_{\lambda_\textrm{\tiny i}}$ 
are rather lengthy to compute because even for the non-interacting case $\Gamma$ takes a non-trivial 
form (in contrast to $\Phi$). The explicit calculations can be found in Ref.~\cite{Kem13}; note 
that in that work the moments $\overline{\Gamma}{}^{(n)}$ are not defined as functional derivatives 
but rather only implicitly [namely via 
Eq.~(14) 
of that work]. Thus, the $\overline{\Gamma}{}^{(n)}$ cannot be expected to obey all the index permutation 
symmetries that functional derivatives would have.\footnote{This implicit definition was dealt 
with consistently in Ref.~\cite{Kem13} but the reader should keep it in mind in order to avoid confusion.} 

Once the flow equations $\dot{\overline{\Gamma}}{}_\lambda^{(n)}$ and the initial conditions $\overline{\Gamma}{}_{\lambda_\ix{i}}^{(n)}$ are determined, a natural truncation is obtained by setting $\overline{\Gamma}{}_{\lambda}^{(n)}=\overline{\Gamma}{}_{\lambda_\ix{i}}^{(n)}$ for all $n \geq m$ for a certain $m$. This is again motivated from a weak coupling perspective (cf. Sect.~\ref{sec:C_flow_truncated_flow_eq_equiv_to_scpt}). At the end of the flow, one obtains an approximation to the interacting $\overline{\Gamma}{}_{\lambda}^{(n)}$ for $n<m$. Along with the non-interacting $\overline{\Gamma}{}_{\lambda}^{(n)}$ for $n\geq m$, 
these moments constitute an approximation to the energy functional which is valid ``locally'' 
around its minimum. In this way, DFT-fRG is able to provide an approximation to the interacting energy functional.

We here specify the (first three) flow equations for the quantum anharmonic oscillator.
For a self-contained presentation we stick to the notation introduced above 
up to a redefinition of $\overline{G}_\lambda$  including a factor $\frac{1}{\beta}$ 
(compare to Ref. \cite{Kem13}).
From $\dot{\overline{\Gamma}}{}^{(1)}_\lambda=0$, one obtains a flow equation for $\overline{G}_\lambda$. Instead of providing a flow equation for $\overline{\Gamma}{}_\lambda^{(2)}$, it is equivalently given for $\overline{W}{}_\lambda^{(2)}$. Adopting the definition for $\overline{\Gamma}^{(n)}$ by functional derivatives and additionally performing the zero 
temperature limit (this implies some minor differences to the equations given in Ref.~\cite{Kem13}), one finds
\begin{eqnarray}
\fl \dot{\overline{\Omega}}_\lambda=&\frac{g}{24}\left[\overline{G}_\lambda^2+\int_0^{\infty} \frac{d \nu_1}{\pi} 
\overline{W}{}^{(2)}_\lambda (\nu_1)\right] ,
\\ \fl \dot{\overline{G}}_\lambda=&\frac{g}{12} \overline{W}{}^{(2)}_\lambda (0) 
\left[ -\overline{G}_\lambda+ \int_0^{\infty} \frac{d \nu_1}{2 \pi} \left(\overline{W}{}^{(2)}(\nu_1)\right)^2 \overline{\Gamma}{}^{(3)}_\lambda(\nu_1,-\nu_1) \right] ,
\end{eqnarray}
\begin{eqnarray}
\fl \dot{\overline{W}}{}^{(2)}_\lambda\!(\nu)=-\frac{g}{12} \overline{W}{}^{(2)}_\lambda\!(\nu)\!\!\!\left.\vphantom{\sum}\right.^2\!\left\{1+\frac{12}{g} \dot{\overline{G}}_\lambda \overline{\Gamma}{}^{(3)}_\lambda\!(0,\nu)- \!\int_0^{\infty} \!\frac{d \nu_1}{2\pi} \overline{W}{}^{(2)}_\lambda\!(\nu_1)\!\!\!\left.\vphantom{\sum}\right.^2 \overline{\Gamma}{}^{(4)}_\lambda\!(\nu_1,\!-\nu_1,\nu)\right.
\nonumber 
\\ \fl \; +\! \left.\int_0^{\infty}\!\frac{d \nu_1}{2\pi} \overline{W}{}^{(2)}_\lambda\!(\nu_1)\!\!\!\left.\vphantom{\sum}\right.^2 \!\left[\overline{W}{}^{(2)}_\lambda\!(\nu_1\!+\!\nu) \left(\overline{\Gamma}{}^{(3)}_\lambda\!(\nu_1,\nu)\right)^2\!+\overline{W}{}^{(2)}_\lambda\!(\nu_1\!-\!\nu) \left(\overline{\Gamma}{}^{(3)}_\lambda\!(\nu_1,\nu)\right)^2 \right] \right\} .
\end{eqnarray}
The initial conditions are 
\begin{eqnarray}
 \overline{\Omega}_0=\frac{\omega_\textrm{\scriptsize G}}{2}, \qquad \overline{G}_0=\frac{1}{2\omega_\textrm{\scriptsize G}}, \qquad \overline{W}{}^{(2)}_\lambda(\nu)=\frac{2}{\omega_\textrm{\scriptsize G}} \frac{1}{\nu^2+4 \omega_\textrm{\scriptsize G}^2} ,
\\ \overline{\Gamma}{}^{(3)}_0(\nu_1,\nu_2)=-\omega_\textrm{\scriptsize G}^2 \left(\nu_1^2+\nu_1\nu_2 +\nu_2^2+12 \omega_\textrm{\scriptsize G}^2\right) ,
\end{eqnarray}
\begin{eqnarray}
\fl \overline{\Gamma}{}^{(4)}_\lambda\!(\nu_1,\!\nu_2,\!\nu_3)\!=  & \!\!\! \left[2 \omega_\textrm{\scriptsize G}^3\frac{\left(\left(\nu_1\!+\!\nu_2\right)^2 \!+\!\left(\nu_1\!+\!\nu_2\right)\nu_3 \!+\!\nu_3^2\!+\!12 \omega_\textrm{\scriptsize G}^2\right)\left(\nu_1^2\!+\!\nu_1\nu_2\!+\!\nu_2^2\!+\!12\omega_\textrm{\scriptsize G}^2\right)}{\left(\nu_1+\nu_2\right)^2+4\omega_\textrm{\scriptsize G}^2}\right.
\nonumber 
\\ \fl & \hphantom{xxx} \left. - \frac{\omega_\textrm{\scriptsize G}^3 f(\nu_1,\nu_2,\nu_3)}{\left(\left(\nu_1+\nu_2\right)^2+4\omega_\textrm{\scriptsize G}^2\right)\left(\left(\nu_2+\nu_3\right)^2+4\omega_\textrm{\scriptsize G}^2\right)}\right]
\nonumber 
\\ \fl & \hphantom{xx} +\left(\nu_1 \to \nu_2 \to \nu_3 \to \nu_1\right)+\left(\nu_1 \to \nu_3 \to \nu_2 \to \nu_1\right) \vphantom{\frac{\left(f^2\right)}{\left(f^2\right)}},
\end{eqnarray}
with
\begin{eqnarray}
\nonumber \fl f(\nu_1,\nu_2,\nu_3) =&640\omega_\textrm{\scriptsize G}^6 + 48 \omega_\textrm{\scriptsize G}^4 \left[3 \nu_1^{2}+4\nu_2^2+3\nu_3^2+4\nu_1\nu_2+2\nu_1\nu_3+4\nu_2\nu_3\right]
\\ \fl \nonumber &\!+\!4\omega_\textrm{\scriptsize G}^2\! \left[ 3\nu_1^4 +2\nu_2^4 +4\nu_2^3\nu_3+10\nu_2^2\nu_3^2+8\nu_2\nu_3^3+3\nu_3^4+4\nu_1^3\left(2\nu_2+\nu_3\right)\right.
\\ \fl \nonumber &\hphantom{+4\omega^2}\left.+2\nu_1\left(2\nu_2+\nu_3\right)\left(\nu_2^2+\nu_2\nu_3+2\nu_3^2\right)+2\nu_1^2\left(5\nu_2^2+5\nu_2\nu_3+2\nu_3^2\right)\right]
\\ \fl \nonumber &\!+\!\nu_2^2\nu_3^2\left(\nu_2+\nu_3\right)^2+\nu_1\nu_2\nu_3^2\left(\nu_2+\nu_3\right)\left(2\nu_2+\nu_3\right)
\\ \fl \nonumber &\!+\!\nu_1^2\left(\nu_2^4+2\nu_2^3\nu_3+6\nu_2^2\nu_3^2+5\nu_2\nu_3^3+\nu_3^4\right)
\\ \fl &\!+\!\nu_1^3\left(2\nu_2+\nu_3\right)\left(\nu_2^2+\nu_2\nu_3+2\nu_3^2\right)+ \nu_1^4\left(\nu_2^2+\nu_2\nu_3+\nu_3^2\right).
\end{eqnarray}

It is a generic feature of the 2PPI approach that, due to the diagonality of the external source $J$, $\overline{G}$ is not frequency dependent;
this is neither due to the simplicity of the model of the oscillator nor to the truncation.
Also, $\overline{W}{}^{(2)}$ depends on only one frequency (instead of three as it would in a 2PI approach). This is a clear advantage in the numerical treatment of the problem if indeed only the accessible diagonal quantities are desired.

Three truncation schemes are investigated in Ref.~\cite{Kem13}. The natural 
one is obtained by setting $\overline{\Gamma}{}^{(n)}_\lambda$ to their initial values $\overline{\Gamma}{}^{(n)}_{\lambda_\ix{i}}$ for $n\geq 3$. A reduced scheme is defined by setting $\overline{\Gamma}{}^{(3),(4)}_\lambda$ to zero. This reduction allows to gain some analytical insights and solve the flow equations 
analytically 
 \cite{Kem13} 
\begin{equation}
\fl e_\textrm{\scriptsize gs}=\omega_\textrm{\scriptsize G}\left(\frac{1}{4} \frac{g/4!}{1+g/4!}+\sqrt{1+g/4!}-1\right), \qquad \varrho=\frac{1}{2 \omega_\textrm{\scriptsize G}} \frac{1}{1+g/4!} .
\end{equation}
A physically motivated suggestion also leads to an improved scheme \cite{Kem13}: At $\lambda=\lambda_\textrm{\scriptsize i}$, the relation $\overline{G}_{\lambda_\textrm{\tiny i}}=\frac{1}{2\omega_\textrm{\tiny G}}$ holds. For $\lambda \neq \lambda_\textrm{\scriptsize i}$, this relation is inverted to define an effective $\lambda$-dependent frequency characterizing the oscillator $\omega_\lambda^\textrm{\scriptsize eff}=\frac{1}{2 \overline{G}_\lambda}$. This $\omega_\lambda^\textrm{\scriptsize eff}$ is then used to calculate $\overline{\Gamma}{}^{(3),(4)}_\lambda$ according to
\begin{equation}
 \overline{\Gamma}{}^{(3),(4)}_\lambda= \left.\overline{\Gamma}{}^{(3),(4)}_{\lambda_\textrm{\tiny i}} \right|_{\omega_\textrm{\tiny G} \to \omega_\lambda^\textrm{\tiny eff}} .
\end{equation}
Even without 
performing  
explicit calculations, one can state two points: 
on the one hand, 
the motivation for this ad-hoc replacement is physically 
comprehensible: it corresponds to finding the 
harmonic 
oscillator that reproduces 
the present density at the given RG scale $\lambda$. In a sense, this is similar to 
the mean-field idea. On the other hand, the replacement is unsystematic
and carries the danger of double counting of diagrams. 
Whether this is indeed the case, remains to be investigated. Such 
a study 
is technically not straightforward to conduct and we 
refrain from pursuing it here. 

\begin{figure}
 \includegraphics[width=0.5\textwidth]{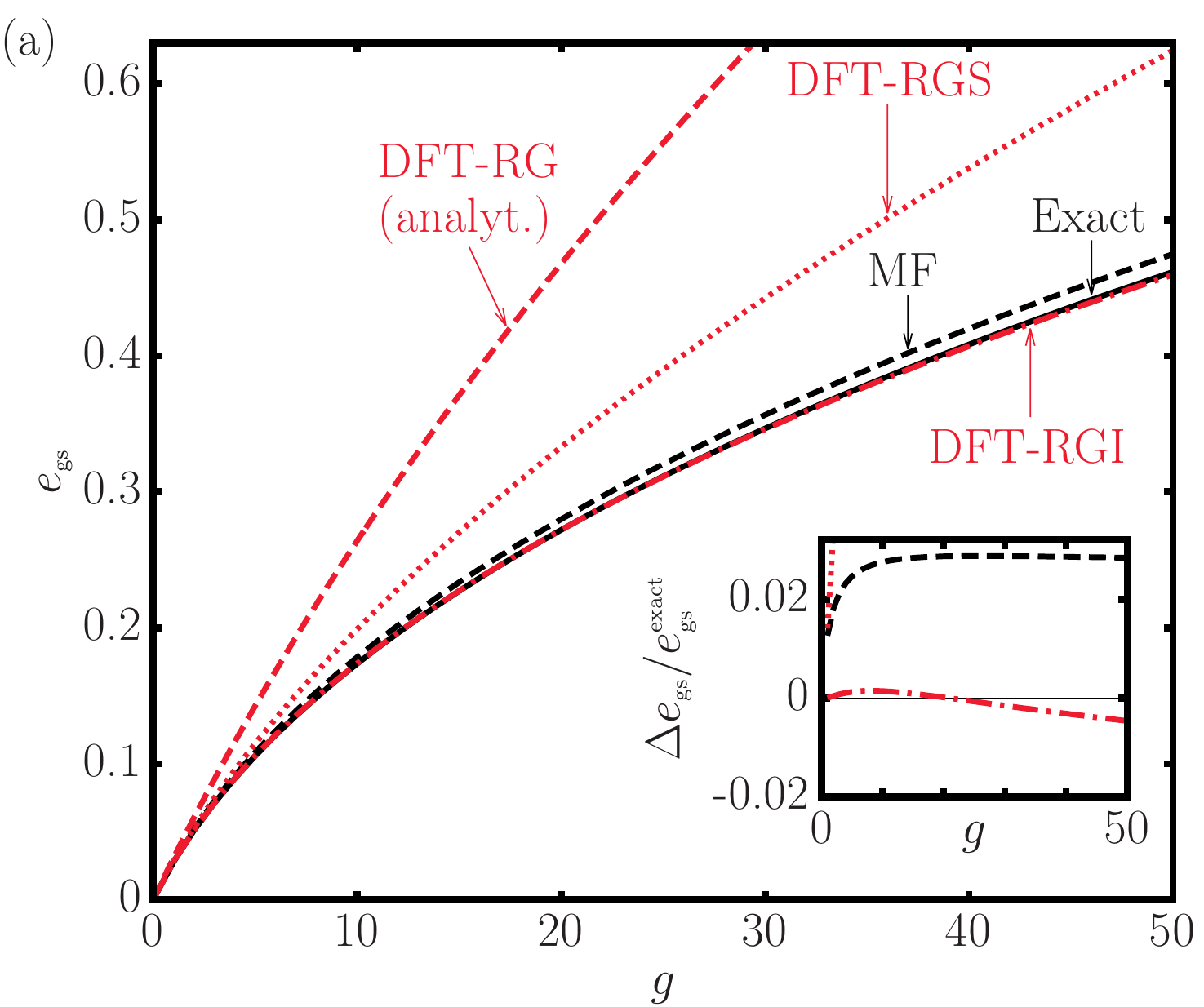}
 \includegraphics[width=0.5\textwidth]{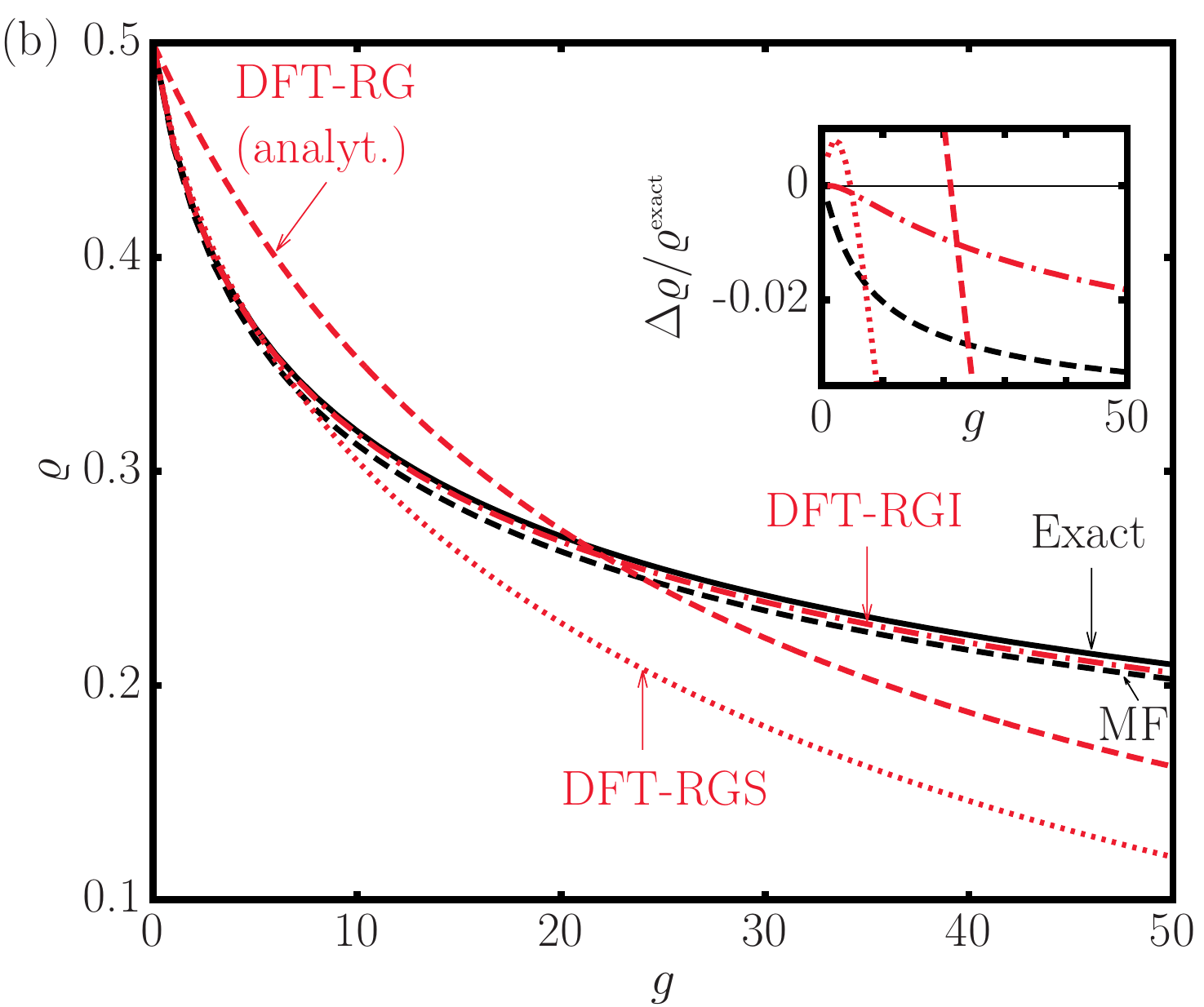}
 \caption{\label{fig:Paper_QAOsc_DFT-RG} 
 Numerical results for (a) the ground state energy and (b) the density as a function of 
 the anharmonicity $g$. 
 Note that this time the main plots show 
 the absolute curves, while the insets show the relative difference to the exact curve. 
 We label the schemes by the same names that were used in Ref.~\cite{Kem13} for better comparability. 
 The natural truncation (DFT-RGS) and improved DFT-fRG (DFT-RGI) data are calculated with 
 $m_\textrm{\scriptsize len}=180$. For comparison, the exact, mean-field (MF) and analytical 
 reduced DFT-fRG (DFT-RG) curves are plotted also.
 }
\end{figure}

Numerical results for the quantum anharmonic oscillator were presented in Ref.~\cite{Kem13} 
for finite temperatures 
$T>0$. 
Our data for the reduced and improved 
DFT-fRG truncation schemes differ from the ones given there.\footnote{The first discrepancy is 
that the free energy curve in the reduced scheme shown in Figure 6 
of Ref.~\cite{Kem13} does not agree with the provided analytical result. The second point involves the numerical data for the free energy and 
density calculated in the improved DFT-fRG scheme shown in Figures 6 and 7 of Ref.~\cite{Kem13}.
The authors of Ref.~\cite{Kem13} informed us that a mistake in their numerical implementation 
is the only origin of the discrepancy to our results.}
As for the other fRG schemes, we here only show temperature $T=0$ curves. 
We numerically solved the DFT-fRG flow equations 
using the frequency grid 
introduced 
in \ref{App:Num_implementation} 
(the reduced scheme has of course an analytical solution). 
The number of flow equations scales $\sim m_\ix{len}$ in the natural as well as in the improved scheme. 
Taking into account the integrations on the right-hand-sides, the effort to calculate the 
set of 
differential equations
in each step of the flow scales $\sim m_\ix{len}^2$. Results 
for all three truncation schemes are shown in Fig.~\ref{fig:Paper_QAOsc_DFT-RG}. 
Note that 
the main plots show 
the ground state energy and the 
density 
as a function of $g$ (and not the 
relative
difference to the exact result as above). This is because the 
reduced and natural DFT-fRG curves are rather far off from the exact curve compared to the 
other RG approaches discussed here. Even for this simple problem of the oscillator, these two 
methods are found to perform much worse than simple mean-field theory. In contrast, the improved 
scheme reproduces the exact curve very well. This is discussed in more depth below.

\section{Comparison of 
the 2PI schemes and conclusion}

\label{sec:comp_schemes}

In this section, the 
anharmonic oscillator results obtained from the 
most successful 
schemes introduced 
above shall be compared to one another, namely the internally closed $C$-flow, modified $U$-flow and 
improved DFT-fRG. Moreover, they shall be set into context by comparing them to results of two 1PI fRG schemes \cite{Hed04}. These two schemes are based on vertex expansion and can be applied to the fermionic many-body problem (cf. Sect.~\ref{sec:part}) in a straight-forward manner - this is in contrast to fRG schemes based on potential expansion.
We additionally give brief accounts of the general advantages and drawbacks of the 
different schemes and on the prospects of applying 
2PI fRG to zero- and one-dimensional fermionic quantum many-body problems.

The first 1PI approach to compare with is the regular Matsubara 
fRG scheme that is obtained by 
truncating the set of flow equations at second order. The second is the 
modification of the first 
suggested by Katanin \cite{Kat04}---it corresponds to replacing the so called single-scale 
propagator with a full $\lambda$-derivative of the full propagator in the flow equation for 
the two-particle vertex. 
The number of flow equations scales $\sim m_\textrm{\scriptsize len}^3$ in 
both 1PI fRG schemes. Both schemes use a sharp imaginary frequency cutoff as flow 
parameter. This means that the integrations on the right-hand-sides of the flow equations 
are ``canceled'' by a $\delta$-function in the regular 1PI fRG whereas this is not 
entirely the case in Katanin 1PI fRG. Consequently, the effort to calculate the 
set of differential equations 
in each step of the flow scales $\sim m_\textrm{\scriptsize len}^3$ 
in regular 1PI fRG and $\sim m_\textrm{\scriptsize len}^4$ in Katanin 1PI fRG. 
For the ground state energy and the density of the anharmonic oscillator, the original numerical 1PI fRG results \cite{Hed04} are used for comparison
to the 2P(P)I schemes.

\begin{figure}
 \includegraphics[width=0.5\textwidth]{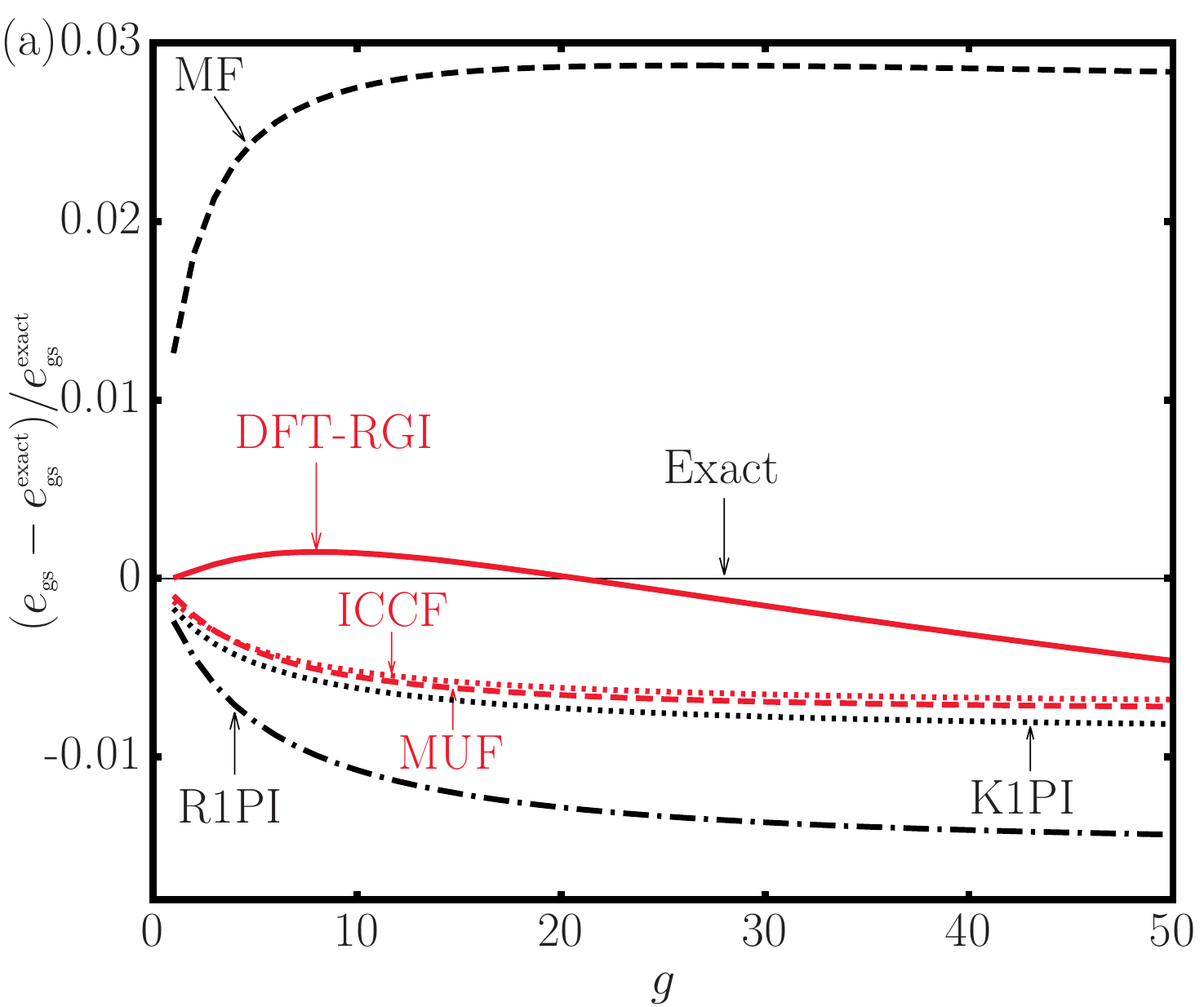}
 \includegraphics[width=0.5\textwidth]{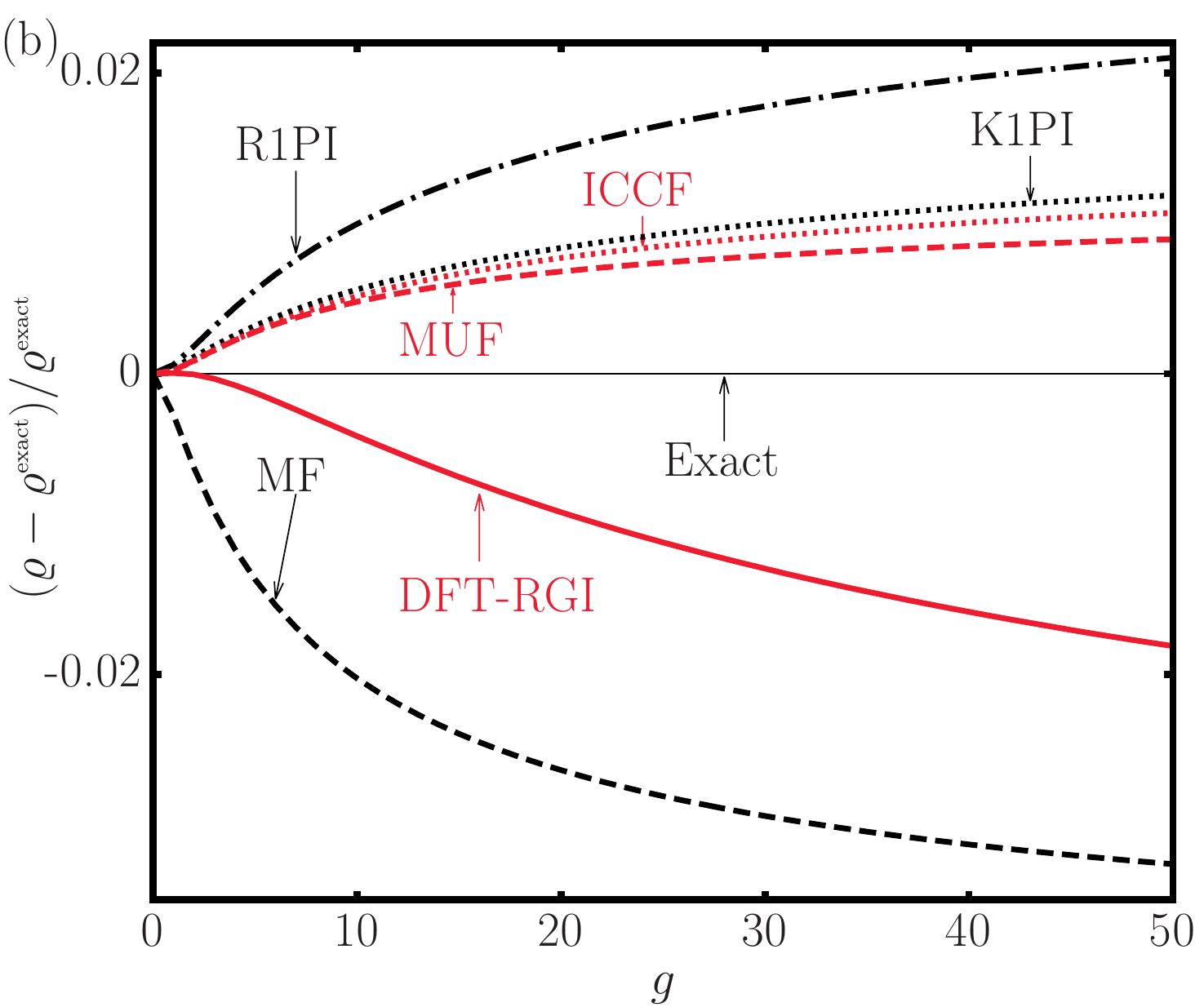}
 \caption{\label{fig:Paper_QAOsc} 
In this figure, 
the numerical results of the most successful 2PI methods for the ground state energy $e_\textrm{\scriptsize gs}$ (a) and the 
density
$\varrho$ (b)
are compared.  
The relative difference to the exact results is shown. 
Also, the mean-field (MF) curve and the regular 1PI fRG (R1PI) and Katanin 1PI fRG (K1PI) curves are 
presented. 
Internally closed $C$-flow (ICCF) is calculated at $m_\textrm{\scriptsize len}=45$, modified 
$U$-flow (MUF) and improved DFT-fRG (DFT-RGI) are calculated at $m_\textrm{\scriptsize len}=180$.}
\end{figure}

Fig.~\ref{fig:Paper_QAOsc} 
shows the numerical results. First of all, all three 2P(P)I schemes outperform the 
1PI 
ones 
(except for improved DFT-fRG in the case of the density for $g \gtrsim 20$). 
That being said, internally closed $C$-flow and modified $U$-flow yield results similar to those of 
Katanin 1PI fRG (but a little better). One should stress, however, that modified $U$-flow does so with 
an effort $\sim m_\textrm{\scriptsize len}^2$ in each step of the flow while the other two methods 
scale 
$\sim m_\textrm{\scriptsize len}^4$. 
We emphasize that these scaling properties also hold for fermionic many-body problems. 

As far as the ground state energy is concerned, improved DFT-fRG 
produces the most accurate results
for the shown anharmonicities $g$. 
Note however, that similar to what is observed for the density around $g=20$, 
the improved DFT-fRG energy presumably becomes less accurate than the 1PI and 2PI ones 
around $g\approx 60$. This is strongly linked to the observation that 
in contrast to the other schemes discussed in this section the improved DFT-fRG 
results for the energy and density deviate in a non-monotonic 
way from the exact one. 
This  behavior might originate from the ad hoc 
replacement $\omega_\textrm{\scriptsize G} \to \omega_\lambda^\textrm{\scriptsize eff}$.
This mean-field-like 
substitution 
is at the same time at the heart of the success of the improved DFT-fRG scheme. This is plausible 
for the quantum anharmonic oscillator, as mean-field produces rather good results for this problem (see above). 
We expect that the improved DFT-fRG scheme is particularly suitable for problems in which 
the mean-field approximation already captures crucial parts of the interaction. 
In DFT-fRG it is generally impossible to access off-diagonal elements of two-point functions. This means that time-non-local properties of propagators cannot be computed although they are needed for the computation of certain observables (e.g. a spectral function). While this is true in equilibrium, time-non-local properties play an even more important role in non-equilibrium situations. Thus, it is not conceivable how to investigate non-equilibrium with this scheme.
We re-emphasize that improved DFT-fRG might be prone to double counting of diagrammatic 
contributions. Prior to any application to fermionic many-body problems this
issue should be thoroughly investigated. 

Also the success of the modified $U$-flow with the Hartree-Fock solution as the starting
point of the RG procedure heavily relies on the accuracy of mean-field theory
for the problem at hand. 
When applied to zero- (quantum dots/impurities) or one-dimensional (quantum wires) fermionic 
many-body problems, modified $U$-flow faces the problem that the unrestricted 
Hartree-Fock starting point shows spurious symmetry breaking. 
This raises the interesting question 
whether the RG procedure can restore the symmetry that is broken in the initial conditions. 
If so, this approximation might turn out to be a favorable combination of the aspects of the physics  
mean-field gets right and those which other fRG schemes (such as e.g. 1PI fRG \cite{Hed04,Kar08,Met12}) 
capture.  
In case the symmetry is not restored by the RG flow, 
one might resort 
to restricted Hartree-Fock as the starting point or employ plain $U$-flow for which the effort 
scales $\sim m_\textrm{\scriptsize len}^2$ in each step of the flow as well. 
The internally closed $C$-flow produces 
rather good results but does not constitute a true advance compared to Katanin 1PI fRG as it also 
scales $\sim m_\textrm{\scriptsize len}^4$ in each step of the flow.

\begin{figure}
\centering
\includegraphics[width=0.5\textwidth]{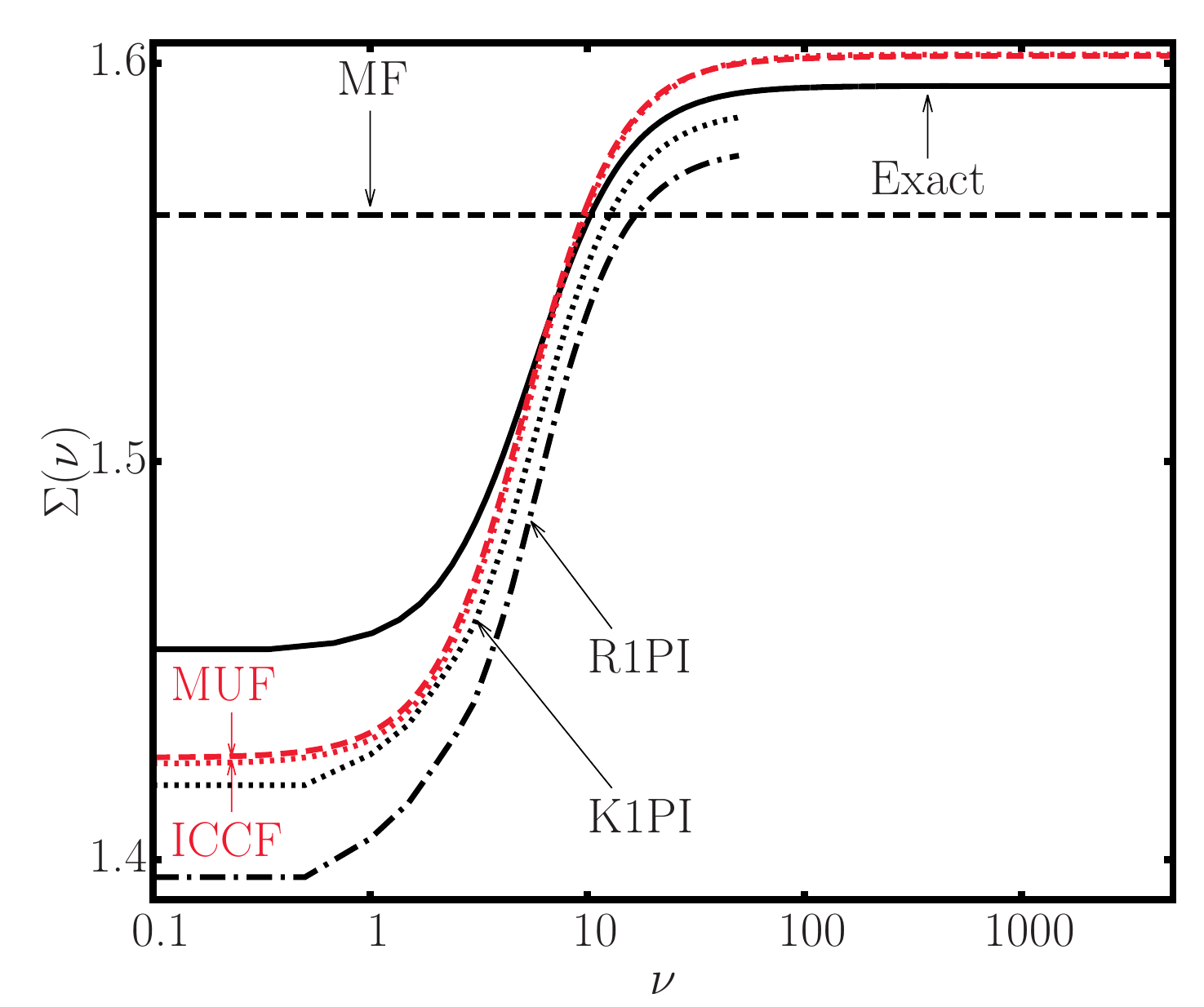}
\caption{\label{fig:Paper_QAOsc_Sigma_freq_dep}
This figure shows the frequency 
dependence of the self-energy (on a log-linear scale) for several approximate 
methods at zero temperature 
and for coupling constant $g=10$---improved DFT-fRG is missing as the self-energy cannot 
reasonably be computed in that scheme. Internally closed $C$-flow (ICCF) is 
calculated at $m_\textrm{\scriptsize len}=45$ and modified $U$-flow (MUF) is calculated at $m_\textrm{\scriptsize len}=180$.}
\end{figure}

As a last point, the frequency dependence of the self-energy 
of the anharmonic oscillator
shall be discussed. 
Investigating  
this quantity promises some deeper understanding of the $e_\textrm{\scriptsize gs}(g)$ and $\varrho(g)$ curves. 
The frequency dependence for fixed interaction $g=10$ 
is shown in Fig.~\ref{fig:Paper_QAOsc_Sigma_freq_dep}.\footnote{As curves for the frequency dependence 
of the self-energy were not shown in the original 
work Ref.~\cite{Hed04} for regular and Katanin 1PI fRG, these curves were calculated within the scope of 
the present one 
using the code of Ref.~\cite{Hed04}. 
An equidistant grid with $\Delta\nu=0.5$ and $m_\ix{len}=100$ was employed.} 
All 
frequency dependent 
fRG 
schemes underestimate $\Sigma(\nu)$ around $\nu=0$; in this regime, the 2PI fRG curves are closer to the exact 
solution.  
Recall that reproducing the correct behavior for small frequencies is more crucial than for large frequencies because $\Sigma(\nu)$ appears in the calculation of the propagator in sum with $\nu^2$ such that the large $\nu$ contribution of $\Sigma(\nu)$  becomes less important. Thus, this observation makes it plausible that the 2PI schemes perform better than the 1PI ones.

\section*{Acknowledgements}
We are grateful to Sandra Kemler and Jens Braun for numerous discussions. We thank the DFG for 
support via the Research Training Group 1995 "Quantum many-body
methods in condensed matter systems".

\section*{Appendix}
\appendix
\section{Numerical implementation}
\label{App:Num_implementation}
In our numerical calculations, we set $\omega_\textrm{\scriptsize G}=1$. The greatest challenge 
in implementing the zero temperature equations is that the frequency axes are continuous and must be 
discretized for the implementation: $\nu \to \nu_n$ (here, $-m_\textrm{\scriptsize len} \leq 
n\leq m_\textrm{\scriptsize len}$ denotes the discretization and not the Matsubara index). We 
require $\nu_{-n}=-\nu_n$. A cubic spline interpolator is used to interpolate function values 
at $\nu \not \in \{\nu_n|-m_\textrm{\scriptsize len} \leq n\leq m_\textrm{\scriptsize len}\}$ and 
to calculate frequency integrals. Almost all right-hand-sides are $\mathcal{O}(\nu_{m_\textrm{\tiny len}}^{-3})$, 
so that $\nu_{m_\textrm{\tiny len}} \approx 5000$ is sufficient to match the accuracy of the differential 
equation solver. Furthermore, it seems advantageous to concentrate the frequencies around zero.

The following grid definition matches the above criteria (let $n\geq0$):
\begin{equation}
 \nu_n = \left\{
 \begin{array}{cl}
  n \Delta\nu &\textrm{for } n\leq m_{\textrm{\scriptsize len},1}
  \\ g_1 +  \frac{f_1^{n-m_{\textrm{\tiny len},1}}-1}{f_1-1} \Delta\nu &\textrm{for } m_{\textrm{\scriptsize len},1} < n \leq m_{\textrm{\scriptsize len},2}
  \\ g_2 f_2^{n-m_{\textrm{\tiny len},2}} &\textrm{for } m_{\textrm{\scriptsize len},2} < n \leq m_{\textrm{\scriptsize len}}
 \end{array}
 \right. .
\end{equation}
Here, $g_1,g_2$ are externally given parameters. $\Delta \nu=g_1/m_{\textrm{\scriptsize len},1}$ where 
integer division defines $m_{\textrm{\scriptsize len},1}=m_\textrm{\scriptsize len}/3$ 
and $m_{\textrm{\scriptsize len},2}=(2 m_\textrm{\scriptsize len})/3$. $f_1$ is determined such 
that $\nu_{m_{\textrm{\tiny len},2}}=g_2$ and $f_2$ is determined such that $\nu_{m_\textrm{\tiny len}}=5120$. 
Preliminary calculations showed that $g_1=2$ and $g_2=20$ are reasonable choices. Numerical convergence with respect 
to $m_\textrm{\scriptsize len}$ was checked carefully.

\section*{References}
\bibliographystyle{unsrt}
\bibliography{paper}

\begin{thebibliography}{10}

\bibitem{Met12}
Walter Metzner, Manfred Salmhofer, Carsten Honerkamp, Volker Meden, and Kurt
  Sch\"onhammer.
\newblock Functional renormalization group approach to correlated fermion
  systems.
\newblock {\em Rev. Mod. Phys.}, 84:299--352, 2012.

\bibitem{Kopietzbook}
Peter Kopietz, Lorenz Bartosch, and Florian Sch\"utz.
\newblock {\em Introduction to the Functional Renormalization Group}.
\newblock Springer, Berlin, 2010.

\bibitem{Pla13}
C.~Platt, W.~Hanke, and R.~Thomale.
\newblock Functional renormalization group for multi-orbital fermi surface
  instabilities.
\newblock {\em Advances in Physics}, 62(4-6):453--562, 2013.

\bibitem{Hed04}
R~Hedden, V~Meden, Th~Pruschke, and K~Sch\"onhammer.
\newblock A functional renormalization group approach to zero-dimensional
  interacting systems.
\newblock {\em Journal of Physics: Condensed Matter}, 16(29):5279, 2004.

\bibitem{Kar06}
C.~Karrasch, T.~Enss, and V.~Meden.
\newblock Functional renormalization group approach to transport through
  correlated quantum dots.
\newblock {\em Phys. Rev. B}, 73:235337, 2006.

\bibitem{Kar08}
C.~Karrasch, R.~Hedden, R.~Peters, Th. Pruschke, K.~Sch\"onhammer, and
  V.~Meden.
\newblock A finite-frequency functional renormalization group approach to the
  single impurity {Anderson} model.
\newblock {\em Journal of Physics: Condensed Matter}, 20(34):345205, 2008.

\bibitem{Hon04}
Carsten Honerkamp, Daniel Rohe, Sabine Andergassen, and Tilman Enss.
\newblock Interaction flow method for many-fermion systems.
\newblock {\em Phys. Rev. B}, 70:235115, 2004.

\bibitem{Kin13}
Michael Kinza, Jutta Ortloff, Johannes Bauer, and Carsten Honerkamp.
\newblock Alternative functional renormalization group approach to the single
  impurity {Anderson} model.
\newblock {\em Phys. Rev. B}, 87:035111, 2013.

\bibitem{Tar14}
C.~Taranto, S.~Andergassen, J.~Bauer, K.~Held, A.~Katanin, W.~Metzner,
  G.~Rohringer, and A.~Toschi.
\newblock From infinite to two dimensions through the functional
  renormalization group.
\newblock {\em Phys. Rev. Lett.}, 112:196402, 2014.

\bibitem{Went14}
N.~Wentzell, C.~Taranto, A.~Katanin, A.~Toschi, and S.~Andergassen.
\newblock Correlated starting points for the functional renormalization group.
\newblock {\em Phys. Rev. B}, 91:045120, Jan 2015.

\bibitem{Str13}
Simon Streib, Aldo Isidori, and Peter Kopietz.
\newblock Solution of the {Anderson} impurity model via the functional
  renormalization group.
\newblock {\em Phys. Rev. B}, 87:201107, 2013.

\bibitem{Jak07}
Severin Jakobs, Volker Meden, and Herbert Schoeller.
\newblock Nonequilibrium functional renormalization group for interacting
  quantum systems.
\newblock {\em Phys. Rev. Lett.}, 99:150603, 2007.

\bibitem{Gez07}
R.~Gezzi, Th. Pruschke, and V.~Meden.
\newblock Functional renormalization group for nonequilibrium quantum many-body
  problems.
\newblock {\em Phys. Rev. B}, 75:045324, 2007.

\bibitem{Jak10b}
Severin~G. Jakobs, Mikhail Pletyukhov, and Herbert Schoeller.
\newblock Nonequilibrium functional renormalization group with
  frequency-dependent vertex function: A study of the single-impurity
  {Anderson} model.
\newblock {\em Phys. Rev. B}, 81:195109, 2010.

\bibitem{Kar10}
C.~Karrasch, M.~Pletyukhov, L.~Borda, and V.~Meden.
\newblock Functional renormalization group study of the interacting resonant
  level model in and out of equilibrium.
\newblock {\em Phys. Rev. B}, 81:125122, 2010.

\bibitem{Ken12}
D.~Kennes, S.~Jakobs, C.~Karrasch, and V.~Meden.
\newblock Renormalization group approach to time-dependent transport through
  correlated quantum dots.
\newblock {\em Phys. Rev. B}, 85:085113, 2012.

\bibitem{Sal04}
Manfred Salmhofer, Carsten Honerkamp, Walter Metzner, and Oliver Lauscher.
\newblock Renormalization group flows into phases with broken symmetry.
\newblock {\em Progress of Theoretical Physics}, 112(6):943--970, 2004.

\bibitem{Ger08}
R.~Gersch, C.~Honerkamp, and W.~Metzner.
\newblock Superconductivity in the attractive {Hubbard} model: functional
  renormalization group analysis.
\newblock {\em New Journal of Physics}, 10(4):045003, 2008.

\bibitem{Ebe13}
Andreas Eberlein and Walter Metzner.
\newblock Effective interactions and fluctuation effects in spin-singlet
  superfluids.
\newblock {\em Phys. Rev. B}, 87:174523, 2013.

\bibitem{Ebe14}
Andreas Eberlein and Walter Metzner.
\newblock Superconductivity in the two-dimensional $t$-$t'$-{Hubbard} model.
\newblock {\em Phys. Rev. B}, 89:035126, 2014.

\bibitem{Mai14}
Stefan~A. Maier, Andreas Eberlein, and Carsten Honerkamp.
\newblock Functional renormalization group for commensurate antiferromagnets:
  Beyond the mean-field picture.
\newblock {\em Phys. Rev. B}, 90:035140, 2014.

\bibitem{Gie02}
Holger Gies and Christof Wetterich.
\newblock Renormalization flow of bound states.
\newblock {\em Phys. Rev. D}, 65:065001, 2002.

\bibitem{Bai04}
Tobias Baier, Eike Bick, and Christof Wetterich.
\newblock Temperature dependence of antiferromagnetic order in the {Hubbard}
  model.
\newblock {\em Phys. Rev. B}, 70:125111, 2004.

\bibitem{Die07}
S.~Diehl, H.~Gies, J.~Pawlowski, and C.~Wetterich.
\newblock Flow equations for the {BCS-BEC} crossover.
\newblock {\em Phys. Rev. A}, 76:021602, 2007.

\bibitem{Fri11}
S.~Friederich, H.~Krahl, and C.~Wetterich.
\newblock Functional renormalization for spontaneous symmetry breaking in the
  {Hubbard} model.
\newblock {\em Phys. Rev. B}, 83:155125, 2011.

\bibitem{Bir05}
Michael~C. Birse, Boris Krippa, Judith~A. McGovern, and Niels~R. Walet.
\newblock Pairing in many-fermion systems: an exact renormalisation group
  treatment.
\newblock {\em Physics Letters B}, 605(3–4):287 -- 294, 2005.

\bibitem{sue06}
Florian Sch\"utz and Peter Kopietz.
\newblock Functional renormalization group with vacuum expectation values and
  spontaneous symmetry breaking.
\newblock {\em Journal of Physics A: Mathematical and General}, 39(25):8205,
  2006.

\bibitem{Hus09}
C.~Husemann and M.~Salmhofer.
\newblock Efficient parametrization of the vertex function, {$\Omega$}--scheme,
  and the $(t,t')$--{Hubbard} model at van {Hove} filling.
\newblock {\em Phys. Rev. B}, 79:195125, 2009.

\bibitem{Gie12}
Kay-Uwe Giering and Manfred Salmhofer.
\newblock Self-energy flows in the two-dimensional repulsive {Hubbard} model.
\newblock {\em Phys. Rev. B}, 86:245122, 2012.

\bibitem{Kat04}
A.~Katanin.
\newblock Fulfillment of {Ward} identities in the functional renormalization
  group approach.
\newblock {\em Phys. Rev. B}, 70:115109, 2004.

\bibitem{Ens05}
T.~{Enss}.
\newblock {\em {Renormalization, Conservation Laws and Transport in Correlated
  Electron Systems}}.
\newblock PhD thesis, Universit\"at Stuttgart, 2005.

\bibitem{Bay61}
Gordon Baym and Leo~P. Kadanoff.
\newblock Conservation laws and correlation functions.
\newblock {\em Phys. Rev.}, 124:287--299, 1961.

\bibitem{Bay62}
Gordon Baym.
\newblock Self-consistent approximations in many-body systems.
\newblock {\em Phys. Rev.}, 127:1391--1401, 1962.

\bibitem{Ves13}
Kambis Veschgini and Manfred Salmhofer.
\newblock {Schwinger}-{Dyson} renormalization group.
\newblock {\em Phys. Rev. B}, 88:155131, 2013.

\bibitem{Dup05}
N.~Dupuis.
\newblock Renormalization group approach to interacting fermion systems in the
  two-particle-irreducible formalism.
\newblock {\em The European Physical Journal B - Condensed Matter and Complex
  Systems}, 48(3):319--338, 2005.

\bibitem{Dup14}
N.~Dupuis.
\newblock Nonperturbative renormalization-group approach to fermion systems in
  the two-particle-irreducible effective action formalism.
\newblock {\em Phys. Rev. B}, 89:035113, 2014.

\bibitem{Wet07}
C.~Wetterich.
\newblock Bosonic effective action for interacting fermions.
\newblock {\em Phys. Rev. B}, 75:085102, 2007.

\bibitem{Sch13}
T.~Sch\"afer, G.~Rohringer, O.~Gunnarsson, S.~Ciuchi, G.~Sangiovanni, and
  A.~Toschi.
\newblock Divergent precursors of the {Mott}-{Hubbard} transition at the
  two-particle level.
\newblock {\em Phys. Rev. Lett.}, 110:246405, 2013.

\bibitem{Jan14}
V.~Jani\ifmmode~\check{s}\else \v{s}\fi{} and V.~Pokorn\'y.
\newblock Critical metal-insulator transition and divergence in a two-particle
  irreducible vertex in disordered and interacting electron systems.
\newblock {\em Phys. Rev. B}, 90:045143, 2014.

\bibitem{Aok02}
K.I. Aoki, A.~Horikoshi, M.~Taniguchi, and H.~Terao.
\newblock {Non-perturbative renormalization group analysis in quantum
  mechanics}.
\newblock {\em Progress of Theoretical Physics}, {108}({3}):{571--590}, 2002.

\bibitem{Gie06}
H.~{Gies}.
\newblock {Introduction to the Functional RG and Applications to Gauge
  Theories}.
\newblock In J.~{Polonyi} and A.~{Schwenk}, editors, {\em Lecture Notes in
  Physics, Berlin Springer Verlag}, volume 852 of {\em Lecture Notes in
  Physics, Berlin Springer Verlag}, page 287, 2012.

\bibitem{Wey06}
Michael Weyrauch.
\newblock Functional renormalization group and quantum tunnelling.
\newblock {\em Journal of Physics A: Mathematical and General}, 39(3):649,
  2006.

\bibitem{Nag11}
S.~Nagy and K.~Sailer.
\newblock Functional renormalization group for quantized anharmonic oscillator.
\newblock {\em Annals of Physics}, 326(8):1839 -- 1876, 2011.

\bibitem{Pol02}
J.~Polonyi and K.~Sailer.
\newblock Effective action and density-functional theory.
\newblock {\em Phys. Rev. B}, 66:155113, 2002.

\bibitem{swe04}
A.~{Schwenk} and J.~{Polonyi}.
\newblock {Towards Density Functional Calculations from Nuclear Forces}.
\newblock {\em arXiv:nucl-th/0403011}, 2004.

\bibitem{Kem13}
Sandra Kemler and Jens Braun.
\newblock Towards a renormalization group approach to density functional theory
  - general formalism and case studies.
\newblock {\em Journal of Physics G: Nuclear and Particle Physics},
  40(8):085105, 2013.

\bibitem{Neg87}
John~W. Negele and Henri Orland.
\newblock {\em Quantum many-particle systems}.
\newblock Addison-Wesley, 1988.

\bibitem{Paw07}
Jan~M. Pawlowski.
\newblock Aspects of the functional renormalisation group.
\newblock {\em Annals of Physics}, 322(12):2831 -- 2915, 2007.

\bibitem{Jak10a}
Severin~G Jakobs, Mikhail Pletyukhov, and Herbert Schoeller.
\newblock Properties of multi-particle {Green}'s and vertex functions within
  {Keldysh} formalism.
\newblock {\em Journal of Physics A: Mathematical and Theoretical},
  43(10):103001, 2010.

\bibitem{Sch05}
Florian Sch\"utz, Lorenz Bartosch, and Peter Kopietz.
\newblock Collective fields in the functional renormalization group for
  fermions, {Ward} identities, and the exact solution of the
  {Tomonaga}-{Luttinger} model.
\newblock {\em Phys. Rev. B}, 72:035107, 2005.

\bibitem{Whi92}
J.~White.
\newblock Self-consistent {Green} functions for the {Anderson} impurity model.
\newblock {\em Phys. Rev. B}, 45:1100--1106, 1992.

\bibitem{Bra12}
Jens Braun.
\newblock Fermion interactions and universal behavior in strongly interacting
  theories.
\newblock {\em Journal of Physics G: Nuclear and Particle Physics},
  39(3):033001, 2012.

\bibitem{Grossbook}
R.M. Dreizler and E.K.U. Gross.
\newblock {\em Density functional theory}.
\newblock Springer, Berlin, 1990.

\end{thebibliography}

\end{document}